\documentclass[a4paper,12pt]{article}

\usepackage{a4}
\usepackage{graphics}
\usepackage{epsfig}
\usepackage{amssymb,amsmath}
\usepackage{color}
\usepackage{cite}
\usepackage{epsf}
\usepackage{wasysym}
\usepackage{subfigure}
\usepackage{rotating,multirow,booktabs}

\definecolor{green3}{rgb}{0.,0.7,0.0}
\definecolor{red1}{rgb}{0.9,0,0}

\def\lsim{\raise0.3ex\hbox{$\;<$\kern-0.75em\raise-1.1ex\hbox{$\sim\;$}}}
\def\gsim{\raise0.3ex\hbox{$\;>$\kern-0.75em\raise-1.1ex\hbox{$\sim\;$}}}

\newcommand{\mneu}[1]{m_{\tilde{\chi}^0_{#1}}}
\newcommand{\neu}[1]{\tilde{\chi}^0_{#1}}
\newcommand{\fb}{\ \mathrm{fb}}
\newcommand{\gev}{\ \mathrm{GeV}}

\begin{document}

\thispagestyle{empty}
\null
\hfill DESY-11-105\\
\null
\vskip .8cm
\begin{center}
{\Large \bf Measurement of CP asymmetries \\[3mm]in neutralino production
		         at the ILC}
\vskip 2.5em

{\large
{    O.~Kittel$^{a,}$\footnote{
        kittel@th.physik.uni-bonn.de},
     G.~Moortgat-Pick$^{b,c,}$\footnote{
       gudrid.moortgat-pick@desy.de},
     K.~Rolbiecki$^{c,}$\footnote{
       krzysztof.rolbiecki@desy.de},
     P.~Schade$^{c,d,}$\footnote{former address, 
       schade.peter@web.de},
     M.~Terwort$^{c,}$\footnote{
       mark.terwort@desy.de}
}}\\[1ex]
{\normalsize \it
$^{a}$ Departamento de F\'isica Te\'orica y del Cosmos and CAFPE,\\
 Universidad de Granada, E-18071 Granada, Spain
}\\
{\normalsize \it
$^{b}$ University of Hamburg, Luruper Chaussee 149, D-22761 Hamburg, Germany
}\\
{\normalsize \it
$^{c}$ DESY, Notkestrasse 85, D-22607 Hamburg, Germany
}\\
{\normalsize \it
$^{d}$ CERN, CH-1211 Geneve 23, Switzerland
}
\vskip 1em
\end{center} \par
\vskip .8cm

\begin{abstract}

  We study the prospects to measure the CP-sensitive triple-product
  asymmetries in neutralino production $e^+\,e^- \to\tilde{\chi}^0_i
  \, \tilde{\chi}^0_1$ and subsequent leptonic two-body decays
  $\tilde\chi^0_i \to \tilde\ell_R \, \ell$, $ \tilde\ell_R \to
  \tilde\chi^0_1 \, \ell$, for $ \ell= e,\mu$, within the Minimal
  Supersymmetric Standard Model. We include a full detector simulation
  of the International Large Detector for the International Linear
  Collider. The simulation was performed at a center-of-mass energy of
  $\sqrt{s}=500$~GeV, including the relevant Standard Model background
  processes, a realistic beam energy spectrum, beam backgrounds and a
  beam polarization of $80\%$ and $-60\%$ for the electron and positron
  beams, respectively. In order to effectively disentangle different
  signal samples and reduce SM and SUSY backgrounds we apply a method
  of kinematic reconstruction. Assuming an integrated luminosity of
  $500~{\rm fb}^{-1}$ collected by the experiment and the performance
  of the current ILD detector, we arrive at a relative measurement
  accuracy of $10\%$ for the CP-sensitive asymmetry in our scenario. We demonstrate that our method of signal selection using kinematic reconstruction can be applied to a broad class of scenarios and it allows disentangling processes with similar kinematic properties.

\end{abstract}

\newpage

\section{Introduction}

Supersymmetry~(SUSY)~\cite{haberkane,Nilles:1983ge} is one of the best
motivated candidates for physics beyond the Standard Model~(SM).
Besides providing a unification of the strong and electroweak gauge
couplings and a suitable cold dark matter candidate, the lightest
stable SUSY particle, SUSY could offer new sources of CP
violation~\cite{Haber:1997if,Ibrahim:2002ry,Ibrahim:2007fb,Ellis:2007kb}.
In the Minimal Supersymmetric Standard Model (MSSM), the complex
parameters are conventionally chosen to be the Higgsino mass parameter
$\mu$, the ${\rm U(1)}$ and ${\rm SU(3)}$ gaugino mass parameters
$M_1$ and $M_3$, respectively, and the trilinear scalar coupling
parameters $A_f$ of the third generation sfermions ($f=b,t,\tau$),
\begin{eqnarray}
\label{eq:phases}
	\mu = |\mu| e^{i \phi_\mu}, \quad  
	M_1 = |M_1| e^{i \phi_{1}}, \quad
	M_3 = |M_3| e^{i \phi_3},   \quad
	A_f = |A_f| e^{i \phi_{A_f}}.
\end{eqnarray}
The sizes of these phases are constrained by experimental bounds from
the electric dipole moments (EDMs). Such experimental limits
generally restrict the CP phases to be small, in particular the phase
$\phi_\mu$~\cite{Barger:2001nu}. However, the extent to which the EDMs
can constrain the SUSY phases depends strongly on the considered model
and its parameters~\cite{Ellis:1982tk,Barger:2001nu,Bartl:2003ju,
  Choi:2004rf,YaserAyazi:2006zw,Ellis:2008zy,Deppisch:2009nj}.

\medskip

Due to cancellations among different contributions to the EDMs, large
CP phases can give CP-violating signals at colliders, as shown for
example in Ref.~\cite{Deppisch:2009nj}. It is important to search for
these signals, since the cancellations could be a consequence of an
unknown underlying structure that correlates the phases. In addition,
the existing EDM bounds could also be fulfilled by including lepton
flavor violating couplings in the slepton
sector~\cite{Bartl:2003ju}.

\medskip

Thus, direct measurements of SUSY CP-sensitive observables are
necessary to determine or constrain the phases independently of EDM
measurements. The phases can change SUSY particle masses, their cross
sections, branching
ratios~\cite{Bartl:2003he,Bartl:2003pd,Bartl:2002uy,Bartl:2002bh,Rolbiecki:2009hk},
and longitudinal polarizations of final
fermions~\cite{Gajdosik:2004ed}. Although such CP-even observables can
be very sensitive to the CP phases, CP-odd (T-odd) observables have to
be measured for a direct evidence of CP violation.

\medskip

CP asymmetries in particle decay chains can be defined with triple
products of final particle
momenta~\cite{Valencia:1994zi,Branco:1999fs}. Due to spin
correlations, such asymmetries show unique hints for CP phases already
at tree level. Thus, triple product asymmetries have been proposed in
many theoretical papers. For the Large Hadron Collider (LHC), triple
product asymmetries have been studied for the decays of
neutralinos~\cite{Bartl:2003ck,Langacker:2007ur,MoortgatPick:2009jy},
stops~\cite{Bartl:2004jr,Ellis:2008hq,Deppisch:2009nj,MoortgatPick:2010wp},
sbottoms~\cite{Bartl:2006hh,Deppisch:2010nc}, and
staus~\cite{Dreiner:2010wj}. In a Monte Carlo~(MC) analysis for stop
decays\cite{Ellis:2008hq,MoortgatPick:2010wp}, it could be shown that
the decay chain can be reconstructed and asymmetries be measured at a
$3\sigma$ level for luminosities of the order of $300~{\rm fb}^{-1}$.
At the International Linear Collider
(ILC)~\cite{Behnke:2007gj,:2007sg,Djouadi:2007ik,TDR} a clearer
identification and a more precise measurement is expected to be
achievable. However, in this context only theoretically-based papers
exist: for instance, neutralino production with
two-~\cite{Bartl:2003tr,Bartl:2003ck,Bartl:2003gr,Choi:2003pq,
  Bartl:2004ut,AguilarSaavedra:2004dz,Choi:2003fs,Bartl:2009pg,Kittel:2004rp}
and three-body
decays~\cite{Kizukuri:1990iy,Choi:1999cc,Bartl:2004jj,AguilarSaavedra:2004hu,
  Choi:2005gt}, charginos with
two-~\cite{Choi:2000ta,Bartl:2004vi,Kittel:2004kd,
  AguilarSaavedra:2004ru,Bartl:2008fu,Dreiner:2010ib} and three-body
decays~\cite{Kizukuri:1993vh,Bartl:2006yv}, also with transversely
polarized beams~\cite{MoortgatPick:2005cw,Bartl:2004xy,Bartl:2005uh,
  Choi:2006vh,Bartl:2006bn,Bartl:2007qy}, have been studied.

\medskip

\begin{figure}[t]
\centering
\includegraphics[scale=1]{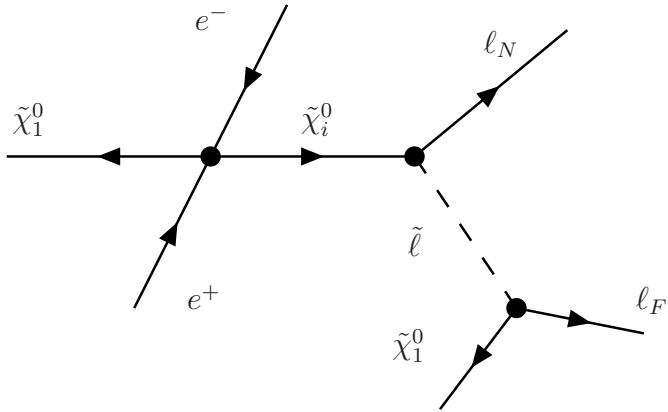}
\caption{\label{shematic picture}
Schematic picture of neutralino production and decay.}
\end{figure}

Therefore, we present in this paper the first experimentally-oriented
analysis with regard to the observation of CP asymmetries on the basis
of a full detector simulation. To show the feasibility of a
measurement of triple product asymmetries, we focus on neutralino
production~\cite{Bartl:2003tr,AguilarSaavedra:2004dz}
\begin{eqnarray}
	e^+ + e^-&\to& \tilde\chi^0_i+\tilde\chi^0_1
  \label{production} 
\end{eqnarray}
with longitudinally polarized beams and the subsequent leptonic
two-body decay of one of the neutralinos into the near lepton
\begin{eqnarray}
   \tilde\chi^0_i&\to& \tilde\ell_R + \ell_N, 
  \label{decay_1} 
\end{eqnarray}
and that of the slepton into a far lepton
\begin{eqnarray}
  \tilde\ell_R&\to&\tilde\chi^0_1+\ell_F; \qquad \ell= e,\mu.
  \label{decay_2}
\end{eqnarray}
Fig.~\ref{shematic picture} shows a schematic picture of neutralino
production and its decay chain.
The CP-sensitive spin correlations of the neutralino in its production
process allow us to probe the phase of the Higgsino mass parameter
$\mu$, and the gaugino parameter~$M_1$.

\medskip

In order to effectively disentangle different signal samples and
reduce SM and SUSY backgrounds we apply a method of kinematic
reconstruction. A similar approach has been studied successfully for
the LHC~\cite{MoortgatPick:2010wp}. The kinematic reconstruction was
also considered to study selectron and neutralino properties at the
ILC~\cite{AguilarSaavedra:2003hw,AguilarSaavedra:2003ur,Berggren:2005gp}.
Here, we demonstrate that it can be used as an effective signal
selection method, greatly improving the sensitivity to the effects of
CP violation. In particular, compared to the previous studies of
process~\eqref{production}, Ref.~\cite{AguilarSaavedra:2004dz}, we are
able to suppress slepton and $WW$ contamination to $\mathcal{O}(10\%)$
level.

\medskip

To investigate in detail the prospect of measuring CP-sensitive
observables at the ILC we perform a full detector simulation of the
International Large Detector (ILD) concept. We include all relevant SM
and SUSY background processes in our study, simulated with a realistic
beam energy spectrum (beamstrahlung and initial state radiation (ISR)), beam backgrounds and a beam polarization of
$(P_{e^-}, P_{e^+}) = (0.8,-0.6)$, which enhances the cross section of
our signal and the size of the asymmetry. We apply the method of
kinematic reconstruction to a preselected sample of signal event
candidates in order to efficiently reject any background and to
disentangle the decays of $\tilde{\chi}^0_2$ from $\tilde{\chi}^0_3$.
We determine the CP asymmetries with the selected signal events and
study the sensitivity to determine the values of the CP phases via a
fit.

\medskip

The paper is organized as follows. In Sec.~\ref{theo frame} we
introduce the theoretical framework for the used CP-sensitive
observable and we apply it for the studied process in the chosen
benchmark scenario. Section~\ref{sec:kinrecoTheo} discusses the
kinematic aspects of signal versus background selection.
Section~\ref{Numerical results} treats the full detector simulation.
Finally in Sec.~\ref{fitting-proc} the SUSY parameters including the
CP phases are derived via a fit of the CP-odd asymmetries together
with masses and cross sections. Appendix~\ref{sec:app3} provides
details for the reconstruction of $W$ and $\tilde{\ell}$ production
and App.~\ref{sec:NeutralinoMixing} recapitulates the neutralino
mixing and its parameters.

\section{Theoretical framework\label{theo frame}}
\subsection{CP-odd observables\label{CP-odd observables}}

CP-violating observables in collider-based experiments are based on
the invariance under CPT$_{\rm N}$, where C is charge conjugation, P
stands for parity transformation and T for time reversal. The index
N denotes 'naive' time reversal, i.e. time reversal but without
interchanging initial with final states and therefore can be tested in
collider-based experiments. At tree level of perturbation theory,
observables odd under T$_{\rm N}$ transformations are also odd under the
'true' time reversal T.

Therefore, it is useful to categorize CP-violating observables in two
classes~\cite{Atwood:2000tu}: those that are even under T$_{\rm N}$
and those that are odd under T$_{\rm N}$ operation. Under the absence
of final state interactions (FSI), CPT$_{\rm N}$-even operators relate
T$_{\rm N}$-odd symmetries uniquely with CP-odd
transformations\cite{Valencia:1994zi}. Contrary, CPT$_{\rm N}$-odd
operators (i.e.\ CP-odd but T$_{\rm N}$-even) can have nonzero
expectation values only if FSI are present that give a non-trivial
phase (absorptive phase) to the amplitude. Such a phase can arise for
instance in loop diagrams.

Examples for T$_{\rm N}$-odd observables are triple products that
arise from the terms $\epsilon[p_1,p_2,p_3,p_4]$, where $p_i$ are
4-vectors representing spins or momenta and $\epsilon$ is the
antisymmetric Levi-Civita tensor. Consequently, such T$_{\rm N}$-odd
signals can only be observed in processes where at least four
independent momenta (or their spin orientations) are involved. The
$\epsilon$-tensor can then be expanded in a series of four triple
products $\epsilon[p_1,p_2,p_3,p_4]=E_1\, \mathbf{p}_2 \cdot
(\mathbf{p}_3\times \mathbf{p}_4)\pm \ldots$ that can be evaluated in
a suitable and specific kinematical system. The T$_{\rm N}$-odd
asymmetries are then composed by the corresponding triple products.

\subsection{Neutralino production and decay processes \label{CP-neutralino}}

Neutralinos are mixed states of the supersymmetric partners of the
neutral gauge and Higgs bosons and depend on the phases $\phi_1$ and
$\phi_{\mu}$, see App.~\ref{sec:NeutralinoMixing}. CP-violating
effects in the neutralino production and decay arise at tree level and
can lead to CP-sensitive asymmetries due to neutralino spin
correlations.

In neutralino production effects from CP-violating phases can only
occur if two different neutralinos are produced, $e^+ e^- \to
\tilde\chi^0_i\tilde\chi^0_j$, $i\neq j$. Each of the produced
neutralinos has a polarization with a component normal to the
production plane~\cite{Choi:1999cc,Choi:2001ww,MoortgatPick:1999di}.
This polarization leads to asymmetries in the angular distributions of
the decay products.

In our process the only T$_{\rm N}$-odd contribution originates from
the production process. It is proportional to $\epsilon[p_{e^+}
p_{e^-} s_{\tilde{\chi}^0_i} p_{\tilde{\chi}^0_i}]$ leading to
$\epsilon[p_{e^+} p_{e^-} p_{{\ell}_N} p_{\tilde{\ell}_R}]$ due to
spin correlations caused by the mentioned neutralino polarization
normal to the production plane.  Applying momentum conservation
$p_{\tilde{\ell}_R}=p_{\tilde{\chi}^0_1}+p_{\ell_F}$ allows one to
extract the T$_{\rm N}$-odd triple product of the beam and the final lepton
momenta~\cite{Bartl:2003tr},
 \begin{eqnarray}
{\mathcal T} &=& 
        ({\mathbf p}_{e^-} \times {\mathbf p}_{\ell_N^+}) \cdot {\mathbf p}_{\ell_F^-}
       \; \equiv \;
  ({\mathbf p}_{e^-},{\mathbf p}_{\ell_N^+}, {\mathbf p}_{\ell_F^-}),
	\label{AT}
\end{eqnarray}
which projects out the CP-sensitive parts. 
The corresponding T$_{\rm N}$-odd asymmetry is then
\begin{eqnarray}
  {\mathcal A}({\mathcal T} )  &=&  
        \frac{\sigma({\mathcal T}>0) - \sigma({\mathcal T}<0)}
             {\sigma({\mathcal T}>0) + \sigma({\mathcal T}<0)},
\label{eq:asyth}
\end{eqnarray}
where $\sigma$ is the cross section for neutralino production and
decay, Eqs.~(\ref{production})-(\ref{decay_2}).
Note: in the case of the 3-body neutralino decay, $\tilde{\chi}^0_i\to
\ell^+\ell^-\tilde{\chi}^0_1$, one also obtains T$_{\rm N}$-odd
contributions originating only from the decay
process~\cite{Choi:2005gt,Ellis:2008hq}. Therefore further T$_{\rm
  N}$-odd asymmetries can be composed that contribute also in case of
same-neutralino pair production and can offer tools for disentangling
the different phases.
The CP-sensitive asymmetries, Eq.~\eqref{eq:asyth}, depend on the
charge of the leptons~\cite{Deppisch:2009nj} and the following
relations are given:
\begin{eqnarray} \label{eq:leptonexch}
     {\mathcal A}({\mathbf p}_{e^-}, {\mathbf p}_{\ell_N^+}, {\mathbf p}_{\ell_F^-})
&=& -{\mathcal A}({\mathbf p}_{e^-}, {\mathbf p}_{\ell_N^-}, {\mathbf p}_{\ell_F^+})
   \nonumber \\
&=& -{\mathcal A}({\mathbf p}_{e^-}, {\mathbf p}_{\ell_F^-}, {\mathbf p}_{\ell_N^+})
  \nonumber \\
& =&  +{\mathcal A}({\mathbf p}_{e^-}, {\mathbf p}_{\ell_F^+}, {\mathbf p}_{\ell_N^-}),
\end{eqnarray}
neglecting FSI contributions.
Note that a true CP-odd asymmetry, where also an absorptive phase from FSI or 
finite-width effects is automatically eliminated,
can be defined as
\begin{eqnarray}\label{eq:genuinecp}
{\mathcal A}^{\rm CP} &=& \frac{1}{2}({\mathcal A} -\bar{\mathcal A} ),
\end{eqnarray}
where $\bar{\mathcal A}$ denotes the T$_{\rm N}$-odd asymmetry for the
CP-conjugated process~\cite{Jarlskog:1989bm}. In our case this leads
to a separate measurement of asymmetries for a positive ($\ell^+_N$)
and negative ($\ell^-_N$) near lepton. If ${\mathcal A}^{\rm CP} \neq
0$ holds than we observe a genuine CP-violating effect. Therefore, it
is important to tag the charge of the near and far leptons in order to
establish CP violation in the process
\eqref{production}-\eqref{decay_2}.

\medskip

\begin{figure}[t]
  \centering
\subfigure[]{ \hspace{-1.3cm} \includegraphics[width=0.44\textwidth]{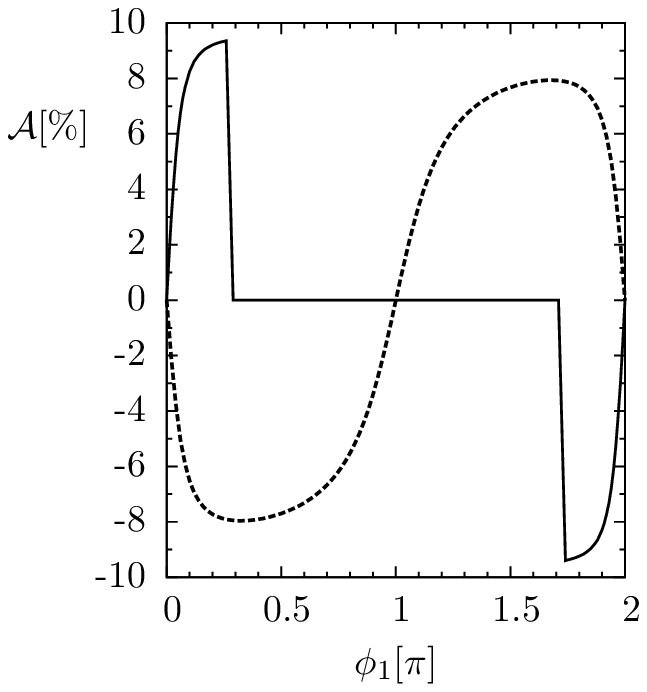}
}
\hspace{1.5cm}
\subfigure[]{ \hspace{-1.3cm} \includegraphics[width=0.44\textwidth]{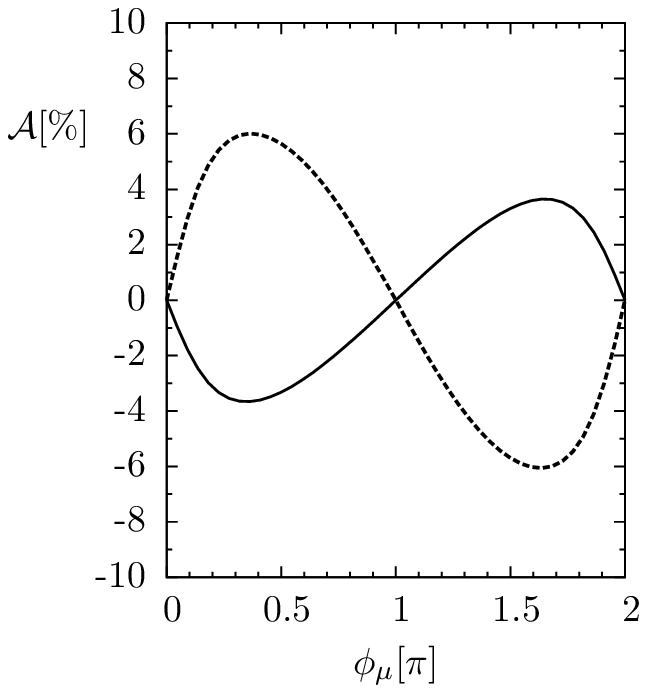}}
  \caption{
       Dependence of the asymmetry ${\mathcal A}$,   Eq.~\eqref{eq:asyth},
       (a) on the phase  $\phi_1$ (with $\phi_\mu=0$),
      (b) the  phase  $\phi_\mu$ (with $\phi_1=0$),
     for neutralino production $e^+e^- \to \tilde\chi^0_1\tilde\chi^0_{2}$
      (solid),  and
      $e^+e^- \to \tilde\chi^0_1\tilde\chi^0_{3}$  (dashed),
        and subsequent decay 
       $\tilde\chi_i^0\to\tilde\ell_{R} \ell_N$,
       $\tilde\ell_{R}  \to \tilde\chi_1^0 \ell_F$,
      (for $\ell=e$ or  $\mu$), at $\sqrt s =500$~GeV and 
          polarized beams $(P_{e^-}, P_{e^+}) = (0.8,-0.6)$.
       The other MSSM parameters are given in Tab.~\ref{tab:scenario}.
     In the left panel, along the flat line of the asymmetry (solid) 
    the decay $\tilde\chi^0_{2}\to \tilde\ell_{R} \ell$ is closed.
}
\label{fig:Asym}
\end{figure}

\subsection{Benchmark scenario}\label{sec:benchmark}

\begin{table}[t]
\renewcommand{\arraystretch}{1.7}
\vspace{1cm}
\begin{center}
\begin{tabular}{cccccccc} \toprule
 
  \multicolumn{1}{c}{$M_2$} 
& \multicolumn{1}{c}{$|M_1|$}
& \multicolumn{1}{c}{$|\mu|$}
& \multicolumn{1}{c}{$\phi_{1}$}  
& \multicolumn{1}{c}{$\phi_\mu$}  
& \multicolumn{1}{c}{$\tan{\beta} $}
& \multicolumn{1}{c}{$M_{\tilde{E}}$}
& \multicolumn{1}{c}{$M_{\tilde{L}}$}
\\\hline
   \multicolumn{1}{c}{$300~{\rm GeV}$} 
&  \multicolumn{1}{c}{$150~{\rm GeV}$} 
& \multicolumn{1}{c}{$165~{\rm GeV}$}
& \multicolumn{1}{c}{$0.2\pi$}  
& \multicolumn{1}{c}{$0$} 
& \multicolumn{1}{c}{$10$}
& \multicolumn{1}{c}{$166~{\rm GeV}$}
& \multicolumn{1}{c}{$280~{\rm GeV}$}
\\\bottomrule
\end{tabular}
\end{center}
\renewcommand{\arraystretch}{1.0}
\caption{
MSSM parameters of the benchmark scenario at the electroweak scale, see Sec.~\ref{sec:benchmark}.
\label{tab:scenario}}
\end{table}

For our full simulation study, we have chosen a benchmark scenario
with the relevant MSSM parameters given in Tab.~\ref{tab:scenario}.
Since the phase of the Higgsino mass parameter is strongly constrained
by EDM bounds, we have set it to zero. The value of the gaugino phase
$\phi_1=0.2\pi$ approximately corresponds to the maximum of the CP
asymmetries, see Fig.~\ref{fig:Asym}.
The scenario was chosen to have an enhanced neutralino mixing close to
a level-crossing of the neutralino states $\tilde\chi_2^0$ and
$\tilde\chi_3^0$ for $\phi_1 = 0$, and of $\tilde\chi_1^0$ and
$\tilde\chi_2^0$ for $\phi_1 = \pi$, which leads to large
CP asymmetries.

\begin{figure}[t]
\vspace{1cm}
 \centering
  \subfigure[]{\label{fig:neu2edist}%
    \includegraphics[scale=1.1]{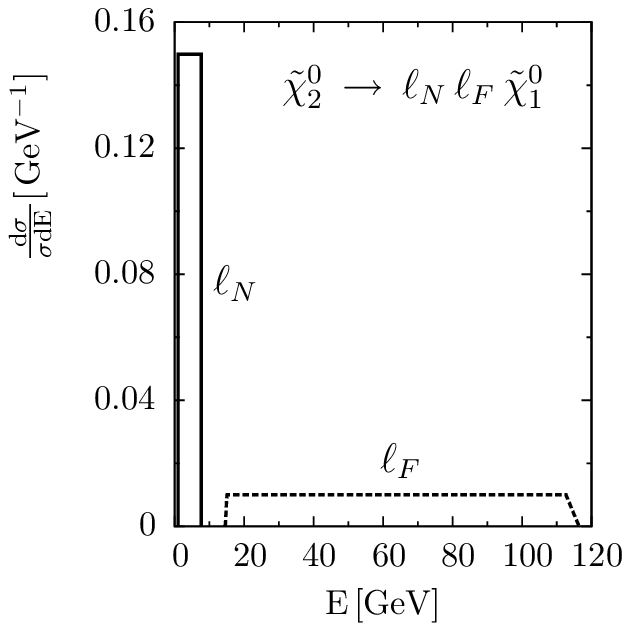}
  } \hspace{0.8cm}
  \subfigure[]{\label{fig:neu3edist}%
    \includegraphics[scale=1.1]{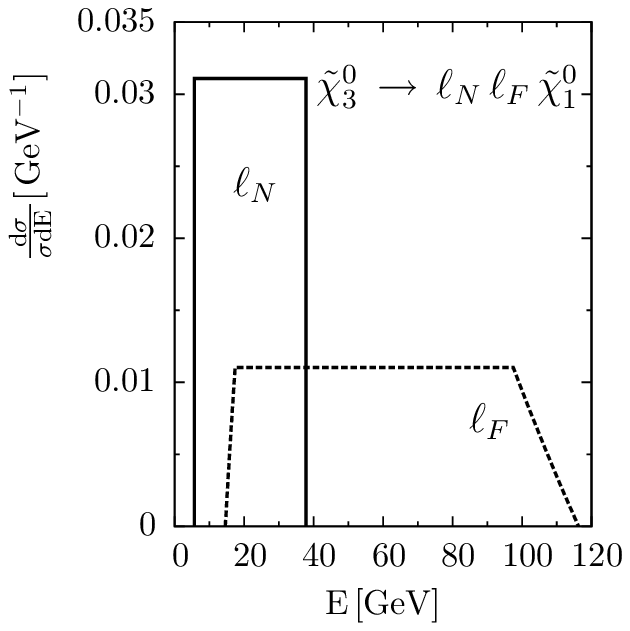}
  }
  \caption{ Energy distributions (each normalized to 1) of the near
    lepton $\ell_N$ (solid), and the far lepton $\ell_F$ (dashed),
    from neutralino production $e^+e^- \to
    \tilde\chi^0_1\tilde\chi^0_i$ for (a)~$i=2$, and (b)~$i=3$, with
    subsequent decay $\tilde\chi_i^0\to \tilde \ell_R \ell_N$,
    $\tilde\ell_R\to \tilde\chi_1^0\ell_F$, for the benchmark scenario
    as given in Tabs.~\ref{tab:scenario} and \ref{tab:masses}.
\label{fig:mu160_edist_plot}
}
\end{figure}

\medskip Further, we have assumed beam polarizations of
$(P_{e^-},P_{e^+})=(0.8,-0.6)$ enhancing slightly the SUSY cross
section and the asymmetries. At the same time, this choice suppresses
the background from $WW$-pair production, $\sigma(e^+e^-\to WW) =
0.7$~pb (compared with 7~pb for unpolarized beams), and also chargino
pair production $\sigma(e^+e^-\to\tilde\chi_i^\pm\tilde\chi_j^\mp) =
110$~fb (410~fb, respectively).
Thus, the scenario is optimized to yield large asymmetries, large
cross sections and sizable neutralino branching ratios into electrons
and muons, as listed in Tab.~\ref{tab:masses}. We have set
$A_{\tau}=-250$ GeV in the stau sector, which has low impact on the
neutralino branching ratios. Also we have chosen the slepton masses
such that $\tilde\chi_2^0$ is close in mass with the slepton
$\tilde\ell_{R}$. This leads to soft leptons from the decay
$\tilde\chi_2^0 \to \tilde\ell_{R} \ell$, as can be seen in
Fig.~\ref{fig:mu160_edist_plot}, which has to be taken care of in the
lepton identification described in Sec.~\ref{sec:preselection}.

\begin{table}[t]
\renewcommand{\arraystretch}{1.7}
\begin{center}
      \begin{tabular}{lcc}
\toprule

 masses & $   m_{\tilde{\chi}^0_1}  =   117~{\rm GeV}$ & $ m_{\tilde \ell_R} = 166~{\rm GeV} $ \\
 & $  m_{\tilde{\chi}^0_2}  =   169~{\rm GeV}$ & $ m_{\tilde \ell_L} = 280~{\rm GeV}$ \\
 & $   m_{\tilde{\chi}^0_3}  =   181~{\rm GeV}$ & $m_{\tilde\tau_1} =   165~{\rm GeV}$ \\
 & $   m_{\tilde{\chi}^0_4}  =   330~{\rm GeV}$ & $m_{\tilde\tau_2} =   280~{\rm GeV}$ \\
 & $ m_{\tilde{\chi}^\pm_1} =   146~{\rm GeV} $ & $  m_{\tilde\nu} =   268~{\rm GeV} $ \\
 & $m_{\tilde{\chi}^\pm_2} =   330~{\rm GeV}$ & $m_{\tilde\nu_\tau} =   268~{\rm GeV}$ \\ \hline
 cross sections & $\sigma(e^+e^-\to\tilde\chi_1^0\tilde\chi_2^0) = 244~{\rm fb} $ & $\sigma(e^+e^-\to\tilde{e}^+_R\tilde{e}^-_R) = 304~{\rm fb}$ \\
 & $\sigma(e^+e^-\to\tilde\chi_1^0\tilde\chi_3^0) = 243~{\rm fb} $ & $\sigma(e^+e^-\to\tilde{\mu}^+_R\tilde{\mu}^-_R) = 97~{\rm fb}$ \\ \hline
 branching ratios & ${\rm BR}(\tilde\chi_2^0\to \tilde \ell_R \ell) = 55 \%$ & ${\rm BR}(\tilde\chi_2^0\to \tilde \tau_1 \tau) = 45 \%$ \\
 & ${\rm BR}(\tilde\chi_3^0\to \tilde \ell_R \ell) = 64 \%$ & ${\rm BR}(\tilde\chi_3^0\to \tilde \tau_1 \tau) = 36 \%$ \\ \hline
 asymmetries & ${\mathcal A}({\mathcal T} )_{\tilde\chi_1^0\tilde\chi_2^0} = -9.2\% $ & ${\mathcal A}({\mathcal T} )_{\tilde\chi_1^0\tilde\chi_3^0} = 7.7\% $ \\  \bottomrule

\end{tabular}
\end{center}
\renewcommand{\arraystretch}{1.0}
\caption{Masses, production cross sections, neutralino branching ratios and asymmetries, Eq.~\eqref{eq:asyth},
         for the benchmark scenario, see Tab.~\ref{tab:scenario}, calculated using the formulas presented in~\cite{Kittel:2004rp}.
         The ILC cross sections are for $\sqrt s =500$~GeV and 
          polarized beams $(P_{e^-}, P_{e^+}) = (0.8,-0.6)$.
         The branching ratios are summed over  
         $\ell=e,\mu$ and both slepton charges.
         \label{tab:masses}}
\end{table}

In general, our analysis is relevant for scenarios
with strong gaugino-higgsino mixing in the neutralino sector, usually leading to sizable asymmetries.
In particular, for $|\mu|\lsim |M_2| \lsim 300$~GeV the asymmetries can reach several
percent, the neutralino pair-production cross sections reach more than
$50$~fb, and the neutralino branching ratios into electrons and muons are
of the order of several $10$\%, see also Ref.~\cite{Bartl:2003tr} for more details. In any case, the selectron and smuon masses should fulfil $m_{\tilde{\chi}_1^0} < m_{\tilde{\ell}_R} < m_{\tilde{\chi}_2^0}$, so that at least one relevant decay channel remains open. Decreasing the selectron mass will result in larger asymmetries and production cross sections.

\section{Kinematic selection of signal and background\label{sec:kinrecoTheo}}

In order to measure the CP asymmetries, Eq.~\eqref{eq:asyth}, we have
to separate the signal lepton pairs originating from
$\tilde\chi_1^0\tilde\chi_2^0$ and $\tilde\chi_1^0\tilde\chi_3^0$
production, respectively.  This is essential, since in our scenario
the corresponding CP asymmetries, $ {\mathcal
  A}(\tilde\chi_1^0\tilde\chi_2^0) = -9.2\% $ and ${\mathcal
  A}(\tilde\chi_1^0\tilde\chi_3^0) = 7.7\% $, have opposite sign.
Large CP asymmetries naturally occur when the neutralinos are mixed
states of gauginos and Higgsinos, which often implies that they are
close in mass. In addition we need an efficient method for background
separation. The CP-even backgrounds will reduce the asymmetries, since
they contribute to the denominator, but cancel out in the numerator of the
asymmetries, see Eq.~\eqref{eq:asyth}. Among the most severe SM and
SUSY background processes are $W$ pair production and slepton pair
production.

\medskip

Owing to a known center-of-mass energy and a well-defined initial
state one may attempt to perform a full kinematic reconstruction of
the events at the ILC. Unlike in the case of the
LHC~\cite{MoortgatPick:2009jy,MoortgatPick:2010wp}, this is already
possible with very short decay chains. Assuming that the masses of
intermediate and invisible particles are known from other
measurements, the full reconstruction can be performed even when only
two particles are visible in the final state.

\medskip

In the following, we extend the method of Ref.~\cite{Bartl:2005uh} to
reconstruct the pair of signal leptons from the neutralino decay,
Eq.~\eqref{production}. We show that even for two neutralino states
that are close in mass, here $\neu{2}$ and $\neu{3}$, the final
leptons can be correctly assigned to their mother particle. A similar
procedure to identify and suppress background from $W$ and slepton
pair production is described in App.~\ref{sec:app3}, see also
Ref.~\cite{Buckley:2007th}. Finally, we discuss how well the kinematic
selection and reconstruction works at the MC level.

\subsection{Kinematic constraints from neutralino production}\label{sec:neurec}

In the center-of-mass system of neutralino pair production 
the momenta and energies are fixed~\cite{MoortgatPick:1999di} :
\begin{eqnarray}
 E_{\neu{i}} =\frac{s+m_{\neu{i}}^2-m_{\neu{j}}^2}{2 \sqrt{s}},\quad
    E_{\neu{j}} =\frac{s+m_{\neu{j}}^2-m_{\neu{i}}^2}{2 \sqrt{s}},\quad
      |{\mathbf p}_{\neu{i,j}}| =\frac{\lambda^{\frac{1}{2}}
             (s,m_{\neu{i}}^2,m_{\neu{j}}^2)}{2 \sqrt{s}}, 
\end{eqnarray}
with the beam energy $E=\sqrt{s}/2$, the neutralino masses
$\mneu{i}, \mneu{j}$, and $\lambda(x,y,z) = x^2+y^2+z^2-2(xy+xz+yz)$.
The neutralino production is followed by the two-body decay chain of
one of the neutralinos $\neu{i}$ via a slepton,
\begin{eqnarray}
   \tilde{\chi}^0_i&\to& \tilde{\ell}_R + \ell_N \to \tilde{\chi}^0_1 + \ell_F + \ell_N, 
    \quad \ell= e,\mu.
\end{eqnarray}
In our signal process, Eqs.~\eqref{production}-\eqref{decay_2}, we
have $\neu{j} = \neu{1}$ and it escapes undetected.

In the following, we assume that the near and far leptons can be
distinguished via their different energy distributions.\footnote{This
  is not needed for the determination of the asymmetry, see
  Eq.~\eqref{eq:leptonexch}, but will be exploited for the event
  selection.} For our scenario, the leptons from $\neu{1}\neu{2}$
production and decay have distinct energy ranges, see
Fig.~\ref{fig:mu160_edist_plot}(a). The leptons from $\neu{1}\neu{3}$
production and decay only have a small overlap in the energy window
$E_\ell\in [18,38] \gev$, see Fig.~\ref{fig:mu160_edist_plot}(b).
Events are discarded if both leptons happen to fall into this energy
range. We now choose a coordinate system such that the measured
momenta are
\begin{eqnarray}
{\mathbf p}_{\ell_N} &=& |{\mathbf p}_{\ell_N}|~ (0,0,1), \label{defl1}\\
{\mathbf p}_{\ell_F} &=& |{\mathbf p}_{\ell_F}| ~
(\sin \theta_{NF},\;0, \;\cos \theta_{NF}),
\quad \theta_{NF} \in[0,\pi],
\label{def2}\label{defl2}
\end{eqnarray}
where $\theta_{NF}$ is the angle between the near and the far leptons.
In order to fully reconstruct the event, the decay angles of the sleptons need to be resolved
\begin{eqnarray}
{\mathbf p}_{\tilde\ell} &=& |{\mathbf p}_{\tilde\ell}|~
  (\sin b \cos B,\; \sin b \sin B, \;\cos b), 
  \quad b\in[0,\pi],\quad B\in[0,2\pi].
  \label{defmomenta}
\end{eqnarray}
The slepton momentum, ${\mathbf p}_{\tilde\ell}^2=
E_{\tilde\ell}^2-m_{\tilde\ell}^2$, is already fixed due to energy
conservation, $E_{\tilde\ell}=E_{\neu{i}}-E_{\ell_N}$. Using also
momentum conservation, $ {\mathbf p}_{\neu{i}}^2 = ({\mathbf
  p}_{\ell_N}+{\mathbf p}_{\tilde\ell})^2, $ the polar angle, $b =
\varangle \, ({\mathbf p}_{\ell_N},{\mathbf p}_{\tilde\ell})$, can be
determined
\begin{eqnarray}
	\cos b &=& \frac{{\mathbf p}_{\neu{i}}^2- {\mathbf p}_{\ell_N}^2-
		{\mathbf p}_{\tilde\ell}^2}{2 |{\mathbf p}_{\ell_N}||{\mathbf
			p}_{\tilde\ell}|}, \qquad
	\sin b = +\sqrt{1-\cos^2 b}.
  \label{eq:cosb}
\end{eqnarray}
Using also momentum conservation in the slepton decay, $ {\mathbf
  p}_{\tilde\ell} = {\mathbf p}_{\ell_F}+{\mathbf p}_{\neu{1}}, $ a
similar relation can be obtained for the azimuthal angle
\begin{eqnarray}
\cos B  &=& 
\frac{1}{ \sin b \sin \theta_{NF}}
\left( 
	\frac{{\mathbf p}_{\ell_F}^2+{\mathbf p}_{\tilde\ell}^2
		-{\mathbf p}_{\neu{1}}^2
	}{2 |{\mathbf p}_{\ell_F}||{\mathbf p}_{\tilde\ell}|}
                  -  \cos b \cos \theta_{NF}
\right),
\label{eq:angleb}
\end{eqnarray}
and ${\mathbf p}_{\neu{1}}^2= E_{\neu{1}}^2-m_{\neu{1}}^2$ is obtained
from energy conservation
$E_{\neu{1}}=E_{\neu{i}}-E_{\ell_N}-E_{\ell_F}$. The kinematic
variables $\cos b$ and $\cos B$ solely depend on the center-of-mass
energy $s$, the lepton energies, $ E_{\ell_N}$ and $E_{\ell_F}$, the
angle between the leptons, ${\mathbf p}_{\ell_N} \cdot {\mathbf
  p}_{\ell_F}$, and finally on the contributing particle masses,
$m_{\tilde\chi_i^0}$, $m_{\tilde\chi_1^0}$, and $m_{\tilde\ell_R}$.
Thus, there only remains an ambiguity for $\sin B$, since for
$B\in[0,2\pi]$ we have $\sin B = \pm\sqrt{1-\cos^2 B}$. This ambiguity
in the azimuthal angle is irrelevant for the efficiency of the event
selection\footnote{ The ambiguity is related to the neutralino
  momentum, for which we have two possible solutions
               \begin{eqnarray}
                  {\mathbf p}_{\neu{i}}={\mathbf p}_{\ell_N}
                   + {\mathbf p}_{\tilde\ell}=
                   \left(\begin{array}{c}
                    \phantom{\pm}
                    |{\mathbf p}_{\tilde\ell}|\sin b \cos B\\
                    \pm|{\mathbf p}_{\tilde\ell}|\sin b \sin B\\
                    |{\mathbf p}_{\ell_N}|+|{\mathbf p}_{\tilde\ell}|\cos b
                  \label{chipm}
           \end{array}\right).
           \end{eqnarray}
           For this reason, the neutralino production plane cannot be resolved
           in $e^+e^-\to\tilde\chi_i^0\tilde\chi_j^0$ processes for $j=1$.
           For $j\ge2$, the decay of $\tilde\chi_j^0$ could be included 
           in order to reconstruct the production plane~\cite{Bartl:2005uh}. 
           In that case larger triple product asymmetries can be studied, 
           which include the neutralino momentum itself~\cite{Bartl:2003tr}.
        }.

\subsection{Method of kinematic event selection}\label{sec:BGSupp}

For a given lepton pair, we apply the following kinematic selection
method. The aim is to assign the correct origin of the lepton pair,
which can be signal, $e^+e^-\to\tilde\chi_i^0\tilde\chi_1^0$, or
background $e^+e^-\to W^+W^-$, $\tilde{\ell}^+_R\tilde{\ell}^-_R$.
Thus, we have four systems of equations, one for each possible
production process. For each candidate event we employ the following
kinematic selection:

\begin{itemize}

\item We apply the reconstruction procedure from
  Sec.~\ref{sec:neurec}, assuming $\neu{1}\neu{2}$ and
  $\neu{1}\neu{3}$ production. Thus, we calculate $\cos b$ and $\cos
  B$, Eqs.~\eqref{eq:cosb} and \eqref{eq:angleb}, with $m_{\neu{i}} =
  m_{\neu{2}}$ ($m_{\neu{i}} = m_{\neu{3}}$) for $\neu{1}\neu{2}$
  ($\neu{1}\neu{3}$) production.

\item We apply the reconstruction procedure from App.~\ref{sec:app3},
  assuming $WW$ and slepton pair production. Thus, we calculate two
  values of $y^2$, Eq.~\eqref{eq:defy}.

\item The event solves the system of equations if
\begin{equation}
	|\cos b| < 1 \qquad \mathrm{and} \qquad |\cos B| < 1 \;, \label{eq:reco_neu}
\end{equation}
when neutralino production has been assumed, and
\begin{equation}
        y^2 > 0 \;, \label{eq:reco_slWW}
\end{equation}
when $W$/slepton production has been assumed.

\item The event is accepted and labeled as coming from a given process
  only if it solves {\it exactly one} out of the four above mentioned
  systems of equations, i.e. it fulfills condition~\eqref{eq:reco_neu}
  for $\neu{1}\neu{2}$ or $\neu{1}\neu{3}$ production, or
  condition~\eqref{eq:reco_slWW} for $W$ or slepton production.

\end{itemize}

In order to demonstrate the efficiency of this procedure we perform a
Monte Carlo simulation of $\neu{1}\neu{2}$, $\neu{1}\neu{3}$,
$W^+W^-$, and $\tilde{\ell}_R^+\tilde{\ell}_R^-$ production and their
leptonic decays, using \texttt{Whizard 1.96}~\cite{Kilian:2007gr}. We
use the MSSM parameters, Tab.~\ref{tab:scenario}, with an integrated
luminosity of ${\mathcal{L}} = 500 \fb^{-1}$, and a beam polarization
of $(P_{e^-}, P_{e^+}) = (0.8,-0.6)$ with realistic beam
spectra\footnote{We include ISR and beamstrahlung, which slightly degrade the number of
  reconstructed events even if the correct process is assumed for a
  given event. Additionally, these effects will increase the number of
  false solutions, leading to wrong assignments.}. In
Tab.~\ref{tab:reconstruction} the results of the event selection are
summarized. Without any additional cuts, the selection method gives an
excellent separation between the different samples at the MC level.
This method is still performing well after a full detector simulation,
as demonstrated in the following section.

\begin{table}[!t] \renewcommand{\arraystretch}{1.3}
\begin{center}
\vspace{0.3cm}
\begin{tabular}{ccccccc} \cmidrule[\heavyrulewidth]{4-7} & & &
  \multicolumn{4}{c}{system solved {\emph{only}}} \\ \cmidrule{4-7} &
  & & $\neu{1}\neu{2}$& $\neu{1}\neu{3}$ & $\tilde{\ell}_R^+
  \tilde{\ell}_R^-$ & $W^+W^-$ \\ \cmidrule[\heavyrulewidth]{1-7}
  \multirow{4}*{\begin{sideways}true
      process\phantom{bb}\end{sideways}} & $64$ k & $\neu{1}\neu{2}$ &
  41566 & 788 & 64 & 856 \\ \cmidrule{2-7} & $74$ k & $\neu{1}\neu{3}$
  & 100 & 25513 & 369 & 873 \\ \cmidrule{2-7} & $200$ k &
  $\tilde{\ell}_R^+ \tilde{\ell}_R^-$ & 181 & 1801 & 43919 & 3400 \\
  \cmidrule{2-7} & $8.8$ k & $W^+W^-$ & 0 & 13 & 37 & 6802 \\
  \bottomrule && purity & 99\% & 91\% & 99\% & 57\% \\ \cmidrule{3-7}
  &&efficiency& 65\% & 34\% & 22\% & 77\% \\ \bottomrule
\end{tabular}
\end{center}
\caption{The numbers of leptonic events from the 
         pair production of neutralinos, sleptons and $W$ bosons,
         with their identification according to the kinematic selection 
         procedure at the generator level, see Sec.~\ref{sec:BGSupp}. The events are simulated 
         for our benchmark scenario, Tab.~\ref{tab:scenario}, 
         with an integrated luminosity of ${\mathcal{L}} = 500 \fb^{-1}$ 
         and beam polarization $(P_{e^-}, P_{e^+}) = (0.8,-0.6)$ for $\sqrt{s}= 500$~GeV. 
         \label{tab:reconstruction} }
\end{table}

The method can be successfully applied also for different particle mass spectra. In our benchmark scenario, due to the small difference between $\tilde{\ell}_R$ and $\tilde{\chi}_2^0$ masses, a separation of the near and far leptons, and of the $\tilde{\chi}^0_2$ and $\tilde{\chi}^0_3$ signals was rather straightforward, see Fig.~\ref{fig:mu160_edist_plot}. However, scenarios with different $m_{\tilde{\chi}_2^0} - m_{\tilde{\ell}_R}$ can turn out to be more demanding. In order to test the applicability of the kinematic reconstruction, we consider a scenario with $m_{\tilde{\ell}_R} = 146$~GeV and the other parameters kept as in the benchmark point. It can be regarded as the worst case scenario since the energies of leptons from neutralino $\tilde{\chi}^0_2$, $\tilde{\chi}^0_3$, and slepton decays are all in the 10 to 80~GeV range. Therefore, in the kinematic reconstruction, one has to take into account two possible assignments of the near and far leptons. This, in principle, could result in a reduction of the efficiency. Nevertheless, in the case of $\tilde{\chi}^0_1 \tilde{\chi}^0_3$ production the efficiency is about 40\%. It drops to 8\% for $\tilde{\chi}^0_1 \tilde{\chi}^0_2$ pairs, since they likely also solve the kinematic on-shell conditions for the $\tilde{\chi}^0_1 \tilde{\chi}^0_3$ production process. The purity, however, remains at about 90\%. In half of the cases, one can also correctly and unambiguously assign near and far leptons.

\section{Full detector simulation study}
	\label{Numerical results}

        The next step of our analysis is passing the generated signal
        events and all relevant SM and SUSY background events through
        a full ILD simulation and event reconstruction. After
        discussing the preselection cuts for the leptonic event
        candidates, we apply the kinematic selection as described in
        the previous section.

\subsection{Detector simulation  and event reconstruction}

For the present study we have performed a full simulation of the ILD
detector designed for the ILC. A detailed description of the detector
concept can be found in Ref.~\cite{Group:2010eu}.
The ILD is a concept under study for a multipurpose particle detector
with a forward-backward symmetric cylindrical geometry. It is designed
for an excellent precision in momentum and energy measurements over a
large solid angle. The tracking system consists of a multi-layer
pixel-vertex detector, surrounded by a system of strip and pixel
detectors and a large volume time projection chamber. The track
finding efficiency is 99.5\% for momenta above 1~GeV and angles down
to $\theta=7^\circ$, while the transverse momentum resolution is
$\delta\left(1/{p}_{\rm T}\right)\sim 2\cdot
10^{-5}$~GeV$^{-1}$. The SiW electromagnetic calorimeter~(ECAL) is
highly segmented with a transverse cell size of 5~mm $\times$ 5~mm and
20 layers. It provides an energy resolution of $(16.6\pm
0.1)/\sqrt{E(\text{GeV})}\oplus (1.1\pm 0.1)\%$ for the measurement of
electrons and photons, and also the steel-scintillator hadronic
calorimeter is highly granular and optimized for Particle Flow
reconstruction. The calorimeters are surrounded by a large
superconducting coil, creating an axial magnetic field of 3.5 Tesla.

For the simulation of the ILD, we use the {\it ILD\_00} detector
model, as implemented in the \texttt{Geant4}-based
\texttt{Mokka}~\cite{Agostinelli:2002hh,Musat:2004sp,MoradeFreitas:2004sq}
package.  We have taken into account all active elements, and also
cables, cooling systems, support structures and dead regions.  We have
used the radiation hard beam calorimeter~(BCAL) to reject forward
$\gamma\gamma$ events at low angles.  In particular the modeling of
the response of the BCAL is relevant for the estimation of the
background from events with activity in the very forward regions.
This background was estimated by tracking electrons to the BCAL and
determining the probability of detection from a map of the expected
energy density from beamstrahlung pairs~\cite{Bechtle:2009em}.

All relevant SM backgrounds\footnote{We consider the final states
  listed in Tab.~\ref{tab:cutflow}.} and SUSY signal and background
events are generated using \texttt{Whiz\-ard}~\cite{Kilian:2007gr},
for $\mathcal L = 500~\rm{fb}^{-1}$ and
$(P_{e^-},P_{e^+})=(0.8,-0.6)$. The \texttt{Whiz\-ard} generator provides an ISR structure function that resums leading soft and collinear logarithms, and hard-collinear terms up to the third order~\cite{Skrzypek:1990qs}. The beamstrahlung is simulated using the \texttt{Circe} package~\cite{Ohl:1996fi}. After the detector simulation the
events are reconstructed with \texttt{MarlinReco}~\cite{Wendt:2007iw}.
We have used the Particle Flow concept, as it is implemented in
\texttt{Pandora}~\cite{Thomson:2007zza}.

\subsection{Backgrounds and event preselection}\label{sec:preselection}

In order to clearly measure the CP-violating effects in the
production of neutralinos, we need to have
a clean sample of signal events.
Otherwise the CP asymmetry would be reduced by the CP-even
backgrounds, which enter in the denominator, see Eq.~\eqref{eq:asyth}. 
We therefore apply a number of preselection cuts listed 
in Tab.~\ref{tab:preselectionCuts},
to reject as much background as possible before applying 
the final selection.

\begin{table}
\begin{center}
\vspace{0.5cm}
\renewcommand{\arraystretch}{1.3}
  \begin{tabular}{ll} \toprule
 initial selection & no significant activity in BCAL \\
                   & number of all tracks $N_{\rm tracks} \leq 7$ \\ \hline
 lepton selection  & $\ell^+\ell^-$ pair with $\ell = e,\mu$     \\
                   & $|\cos\theta|<0.99$, min. energy $E>3$~GeV \\
                   &  lower energetic $\ell$ with $E<18$~GeV, or \\ 
                   & \phantom{x} higher energetic $\ell$ with $E>38$~GeV \\
                   & higher energetic $\ell$ with $E\in[15,150]$~GeV \\
                   & $\theta_{\rm acop}>0.2\pi$, $\theta_{\rm acol}>0.2\pi$ \\ \hline
final preselection & ${p}_{\rm T}^{\rm miss}> 20$~GeV  \\
                   & $E_{\rm vis}< 150$~GeV    \\
                   & $m_{\ell\ell}<55$~GeV \\ \bottomrule
  \end{tabular}
\end{center}
\renewcommand{\arraystretch}{1.0}
\caption{Preselection cuts, see Sec.~\ref{sec:preselection} for details.\label{tab:preselectionCuts}}
\end{table}

\subsubsection{Initial selection}
For efficient electron and muon identification, we apply the following
initial selection on the tracks and clusters reconstructed by the
\texttt{Pandora} Particle Flow algorithm:
\begin{itemize}
\item $\displaystyle{\frac{E_{\rm ECAL}}{E_{\rm tot}}} > 0.6 $ and 
      $\displaystyle{\frac{E_{\rm tot}}{p_{\rm track}}}> 0.9$ for electrons; 
    \item $\displaystyle{\frac{E_{\rm ECAL}}{E_{\rm tot}}} < 0.5$ and
      $\displaystyle{\frac{E_{\rm tot}}{p_{\rm track}}} < 0.3$ (0.8)
      for muons, with energy $E >(<)\,12$~GeV,\footnote{The cut on the
        ratio of the total calorimeter energy and the track momentum
        is relaxed for low-energetic muons, which deposit more energy
        in the calorimeters. This ensures a reasonably high muon
        identification efficiency even for low-energetic muons.}
\end{itemize}
where $E_{\rm ECAL}$ is the energy measured in the electromagnetic
calorimeter, $E_{\rm tot}$ is the total measured energy in the
calorimeters, and $p_{\rm track}$ is the measured track momentum in
the tracking detectors. $E$ is the energy of the Particle Flow object
assigned by \texttt{Pandora}, which is derived from the track momentum
in case a track is present or from the deposited energy in the
calorimeters. We require no significant activity in the BCAL to reject
$\gamma\gamma$ events. We select those events with less than eight
tracks\footnote{Although we expect to have only two isolated leptons
  in our signal events, we do not tighten this cut to avoid removing
  signal events due to overlaid $\gamma\gamma$ events.}, $N_{\rm
  tracks} \leq 7$, which efficiently removes all sorts of hadronic
background.

\subsubsection{Lepton selection}\label{sec:lepselection}

We require that two of the tracks form a pair of opposite-sign
same-flavor leptons $\ell^+ \ell^-$, with $\ell =e$ or $\mu$. Only
electrons or muons are selected with a polar angle $|\cos\theta|<0.99$
and a minimum energy $E>3$~GeV.
There is a large contribution from beam induced
$e^+e^-\rightarrow\gamma\gamma e^+e^-\rightarrow \ell\ell e^+e^-$
($\ell=e,\mu,\tau$) background events\footnote{In this study we only
  consider the class of $\gamma\gamma$ background events where both
  photons from the $e^+$ and $e^-$ beam interact via a virtual
  fermion. Interactions where the photon fluctuates into a
  vector-meson or where the photon is highly virtual and the
  interaction is best described as deep inelastic electron scattering
  on a vector-meson are not considered. Since their transverse
  momentum distributions are narrower, it is less likely that they
  contribute to the overall
  background~\cite{Bambade:2004tq,Bechtle:2009em}.}~\cite{Group:2010eu,Bechtle:2009em}.
The two outgoing beam electrons are high-energetic with a small
scattering angle, while the rest of the event forms a system of low
energy and mass. If the beam remnants escape close to the beam pipe,
and cannot be rejected by a low angle veto, the missing transverse
momentum of the event is limited, such that the remaining leptons are
almost back-to-back in the transverse projection~(T). Thus, we apply a
cut on their acoplanarity angle
\begin{equation}
\theta_{\rm acop}>0.2\pi
\qquad {\rm with} \qquad
\theta_{\rm acop}=\pi - \arccos\left(
        \frac{{\mathbf p}_{\ell^+}^{\rm T} \cdot {\mathbf p}_{\ell^-}^{\rm T}}
             {|{\mathbf p}_{\ell^+}^{\rm T}| \;  |{\mathbf p}_{\ell^-}^{\rm T}|}
    \right),
\end{equation}
where $\theta_{\rm acop} = 0$ for back-to-back ($180^\circ$) events.
Electrons or muons from $\gamma\gamma$ induced $\tau\tau$ events
usually have energies below 10 GeV and can therefore be suppressed by
exploiting that the far leptons from SUSY signal decays usually have
higher energies. To do this and to select the signal lepton pairs
$\ell^+ \ell^-$ from the neutralino $\tilde\chi^0_{2,3}$ decays, we
first use their energy distributions, see
Fig.~\ref{fig:mu160_edist_plot}. We keep events where either the
lower energetic lepton has $E<18$~GeV, or the higher energetic lepton
has $E>38$~GeV. In addition the higher energetic lepton is required to
have an energy $E\in[15,150]$~GeV.
Since the signal lepton pairs $\ell^+\ell^-$ originate from the same
parent neutralino, they follow its direction in first approximation.
The lepton pairs from SM decay processes, and also from slepton pair
decays, tend to be more back-to-back, since the leptons originate from
different mother particles. We therefore apply a cut on the
acollinearity angle between the leptons
\begin{equation}
\theta_{\rm acol} >0.2\pi
\qquad {\rm with} \qquad
\theta_{\rm acol}=\pi - \arccos\left(
              \frac{{\mathbf p}_{\ell^+}\cdot{\mathbf p}_{\ell^-}}
               {|{\mathbf p}_{\ell^+}|\, |{\mathbf p}_{\ell^-}|}\right).
\end{equation}

\begin{figure}[t]
  \centering
  \subfigure[]{\label{fig:p_T_miss_distribution}
    \includegraphics[scale=0.35,angle=0]{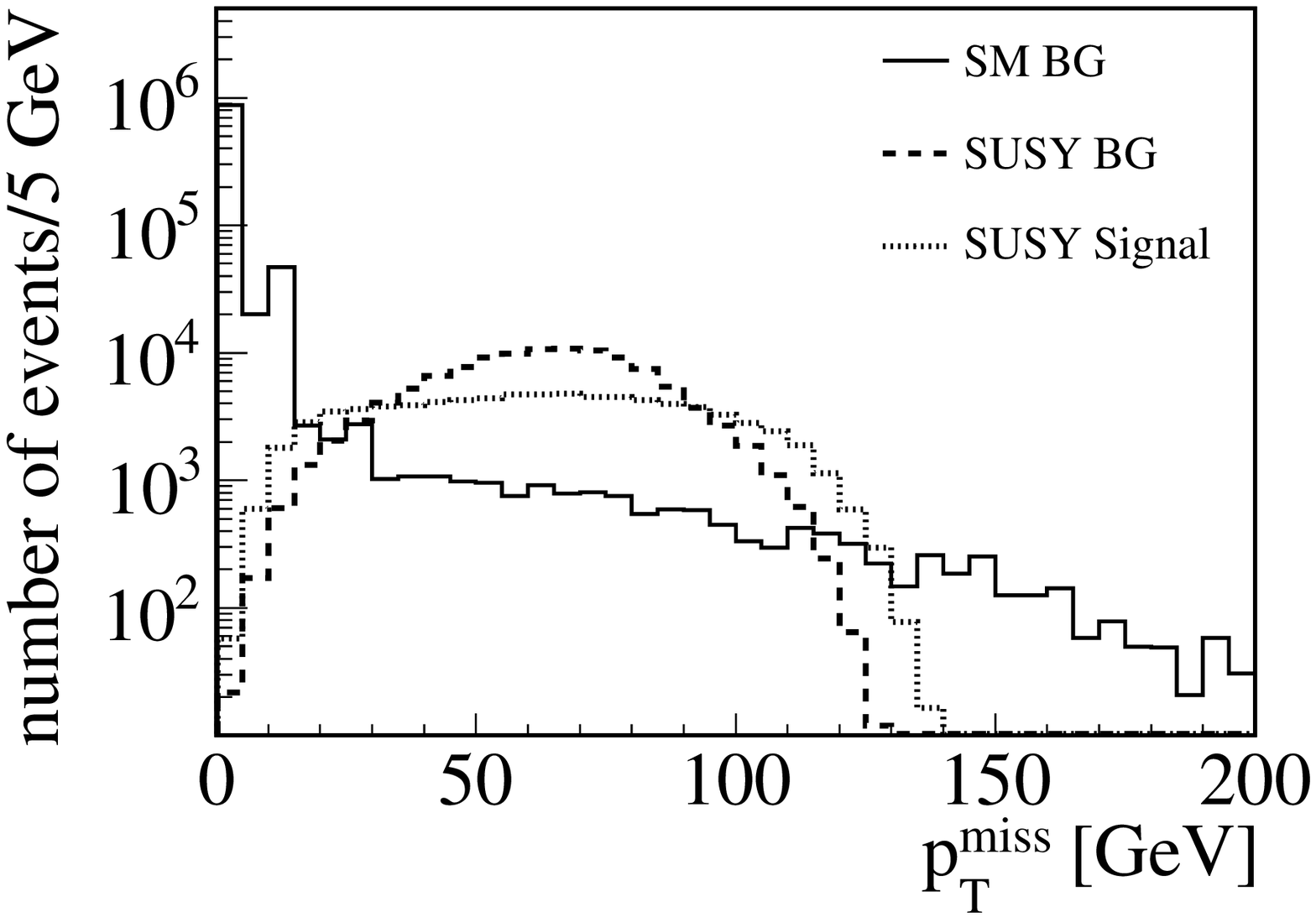}
  }
  \subfigure[]{\label{fig:InvMass_distribution}
    \includegraphics[scale=0.35,angle=0]{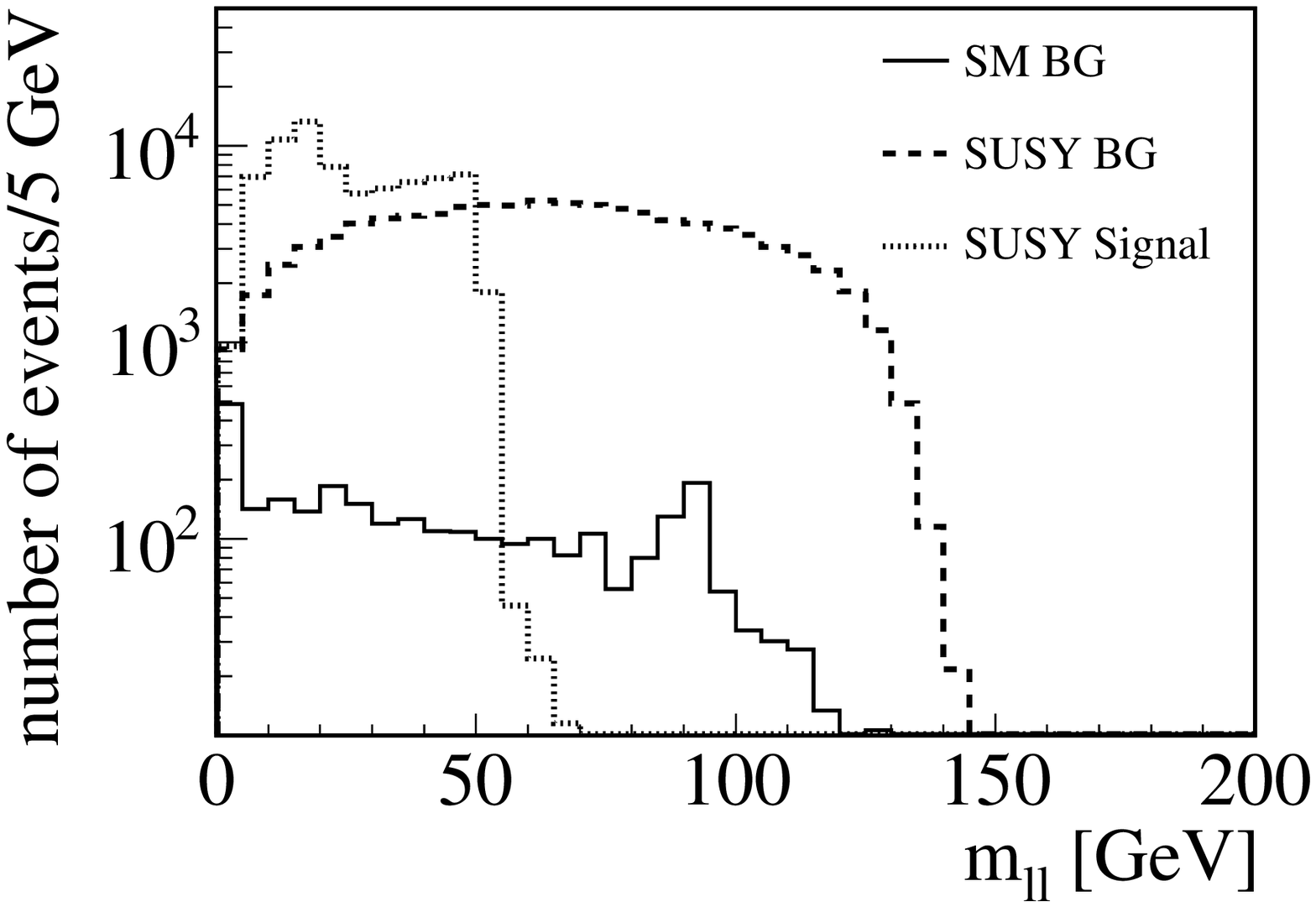}
  }
  \caption{ (a) Missing transverse momentum ${p}_{\rm T}^{\rm
      miss}$ distribution of SM background, SUSY background and SUSY
    signal after the lepton selection, see
    Sec.~\ref{sec:lepselection}. (b) Invariant mass $m_{\ell\ell}$
    distribution of the lepton pair after all preselection cuts except
    the cut on $m_{\ell\ell}$.  The events are simulated for
    ${\mathcal{L}} = 500 \fb^{-1}$, beam polarization $(P_{e^-},
    P_{e^+}) = (0.8,-0.6)$ at $\sqrt{s}= 500$~GeV, and MSSM parameters
    for our benchmark scenario, Tab.~\ref{tab:scenario}.  }
\label{fig:distributions}
\end{figure}

\subsubsection{Final preselection}

In Fig.~\ref{fig:p_T_miss_distribution}, we show the missing
transverse momentum, ${p}_{\rm T}^{\rm miss}$, distribution of
the SM background, the SUSY background, and the SUSY signal after the
lepton selection. The ${\mathbf p}_{\rm T}^{\rm miss}$ is calculated
to balance out the sum of all reconstructed transverse particle
momenta in an event.
Our signal neutralinos $\tilde\chi^0_{2,3}$, but also the background
sleptons, decay into the lightest neutralino, which escapes detection,
thus giving signatures with high ${p}_{\rm T}^{\rm miss}$.
However, most background lepton pairs from beam induced $\gamma\gamma$
events have a transverse momentum typically below $10$~GeV, and are
removed by the cut ${p}_{\rm T}^{\rm miss}>
20$~GeV.\footnote{The spike in the 3rd bin of the ${p}_{\rm
    T}^{\rm miss}$ distribution in
  Fig.~\ref{fig:p_T_miss_distribution} is due to 4
  $\gamma\gamma\rightarrow\ell\ell$ events that have a high event
  weight. Due to limited CPU time and the large cross section of these
  events, it is not possible to simulate an event sample corresponding
  to ${\mathcal{L}} = 500 \fb^{-1}$. The final preselection cuts are
  chosen such that this remaining high cross section background is
  safely removed.} Due to the escaping neutralinos, we also expect a
limited total visible energy $E_{\rm vis}$ in the signal events, and
we apply the cut $E_{\rm vis}< 150$~GeV. The visible energy is
calculated as the sum of all reconstructed particle energies.

Finally we apply a cut $m_{\ell\ell}<55$~GeV on the invariant mass of
the lepton pair, see the distribution in
Fig.~\ref{fig:InvMass_distribution}, after all preselection cuts,
except the cut on $m_{\ell\ell}$. The signal lepton pair from
$\tilde{\chi}^0_3$ ($\tilde{\chi}^0_2$) decays has a sharp endpoint at
$51$~GeV ($22$~GeV), which is also exploited for mass
measurements~\cite{ball,D'Ascenzo:2009zz,Desch:2003vw}. The invariant
mass cut also removes SM backgrounds from $ZZ$ and $WW$ production. In
Fig.~\ref{fig:InvMass_distribution}, we can see the invariant mass
peak of one of the $Z$ bosons decaying into two electrons or muons,
while the other decays into a neutrino pair. The $WW$ events
contribute to the background if they either both decay directly into
same-flavor leptons, or if one of them decays into a $\tau$, which in
turn can complete the same-flavor lepton pair in the event by its
subsequent decay.

The number of remaining events after the lepton selection and the
entire event preselection is listed in Tab.~\ref{tab:cutflow}.  The
most severe remaining SM background stems from $WW$ and $ZZ$
production, while the slepton pair production is the dominant SUSY
background. The difference in the numbers of selected
$\tilde{\chi}^0_2$ and $\tilde{\chi}^0_3$ decays is due to different
cross sections times branching ratios, see Tab.~\ref{tab:masses}, and
due to a reduced muon identification efficiency at low muon momenta,
which reduces the efficiency for the selection of $\tilde{\chi}^0_2$
decays.

\begin{table}
\begin{center}
\vspace{0.5cm}
\renewcommand{\arraystretch}{1.2}
\begin{tabular}{llcc} \toprule class & final state & after lepton
  selection & after preselection \\ \hline signal &
  $\tilde\chi^0_1\tilde\chi^0_2\rightarrow\tilde\chi^0_1\tilde\chi^0_1\ell\ell$
  ($\ell\neq\tau$) & 31543 & 28039 \\
  &
  $\tilde\chi^0_1\tilde\chi^0_3\rightarrow\tilde\chi^0_1\tilde\chi^0_1\ell\ell$
  ($\ell\neq\tau$) & 49084 & 45966 \\ \hline SUSY &
  $\tilde\ell\tilde\ell\rightarrow\tilde\chi^0_1\tilde\chi^0_1\ell\ell$
  ($\ell\neq\tau$)
  & 108302 & 34223 \\
  & $\tilde\chi^0_1\tilde\chi^0_1\tau\tau$ & 5147 & 4076 \\
  & $\tilde\chi^0_1\tilde\chi^0_1\ell\ell\nu\nu$ & 681 & 528 \\ \hline
  SM & $\ell\ell\nu\nu$ & 8241 & 1196 \\
  & $\tau\tau$ & 13017 & 360 \\
  & $\ell\ell$ ($\ell\neq\tau$) & 24113 & 0 \\
  & $qq$  & 1380 & 0 \\
  & $\gamma\gamma$ & 917355 & 272 \\ \bottomrule 
  \end{tabular}
\end{center}
\renewcommand{\arraystretch}{1.0}
  \caption{Number of selected events after lepton selection and final preselection, 
          for ${\mathcal{L}} = 500 \fb^{-1}$, $(P_{e^-}, P_{e^+}) = (0.8,-0.6)$ 
          at $\sqrt{s}= 500$~GeV. The MSSM parameters are given in Tab.~\ref{tab:scenario}.
\label{tab:cutflow}}
\end{table}

\subsection{Signal identification with kinematic event selection
            \label{sec:kinreco}}

          In order to measure our CP
          asymmetry 
          from the preselected events, we now apply the kinematic
          selection procedure, which we have described for the signal
          in Sec.~\ref{sec:kinrecoTheo}, and for the $WW$ and
          $\tilde\ell\tilde\ell$ backgrounds in App.~\ref{sec:app3}.
          The kinematic selection allows us not only to reduce the
          remaining SUSY background from slepton pair production, but
          also to distinguish the lepton pairs which stem from
          $\tilde\chi^0_1\tilde\chi^0_2$ or
          $\tilde\chi^0_1\tilde\chi^0_3$ production and decay. This
          will be essential, since in our benchmark scenario,
          Tab.~\ref{tab:scenario}, the corresponding CP asymmetries
          have roughly equal size, but opposite sign, see Fig.~\ref{fig:Asym}.

\medskip

For each preselected lepton pair, we require that it exclusively
solves only one of the systems of equations, as discussed in
Sec.~\ref{sec:BGSupp}. We reject all other events that solve more than
one system of equations. In Tab.~\ref{tab:reconstruction_results}, we
list the number of preselected events that fulfill this requirement.
For the lepton pairs coming from $\tilde\chi^0_1\tilde\chi^0_2$
decays, the final signal selection efficiency is 29\%, and the total
background contamination of the selected sample is about 8\%. Lepton
pairs from $\tilde\chi^0_1\tilde\chi^0_3$ decays reach a signal
selection efficiency of 27\%, while the total background contamination
of the selected sample is about 16\%.

\begin{table}
\begin{center}
  \label{tab:reconstruction_results}
\vspace{0.5cm}
\renewcommand{\arraystretch}{1.2}
  \begin{tabular}{lcccc} \toprule
    class  & only $\tilde\chi^0_1\tilde\chi^0_2$ & only $\tilde\chi^0_1\tilde\chi^0_3$ & only $\tilde\ell^+_R\tilde\ell^-_R$ & only $W^+W^-$ \\ \hline
    $\tilde\chi^0_1\tilde\chi^0_2\rightarrow\tilde\chi^0_1\tilde\chi^0_1\ell\ell$ ($\ell\neq\tau$) & 18343 & 615 & 51 & 855 \\
    $\tilde\chi^0_1\tilde\chi^0_3\rightarrow\tilde\chi^0_1\tilde\chi^0_1\ell\ell$ ($\ell\neq\tau$) & 290 & 20132 & 372 & 635 \\
    all SUSY background & 1153  & 3055 & 5626   & 951 \\
    all   SM background &   87  &  256 &   44   &  81  \\ \bottomrule
      \hfill     purity & 92\% & 84\% &   -- & --  \\ \hline
      \hfill efficiency & 29\% & 27\% &   --   & -- \\ \bottomrule
  \end{tabular}
\end{center}
\renewcommand{\arraystretch}{1.0}
  \caption{Number of preselected events from Tab.~\ref{tab:cutflow}, 
           that fulfill the requirements of the kinematic selection procedure,
           discussed in Sec.~\ref{sec:kinreco}.}
\end{table}

\subsection{Measurement of the CP asymmetries}\label{sec:asymmetryexp}

\begin{figure}

  \centering
  \subfigure[]{\label{fig:p_T_miss_trip}
    \includegraphics[scale=0.35,angle=0]{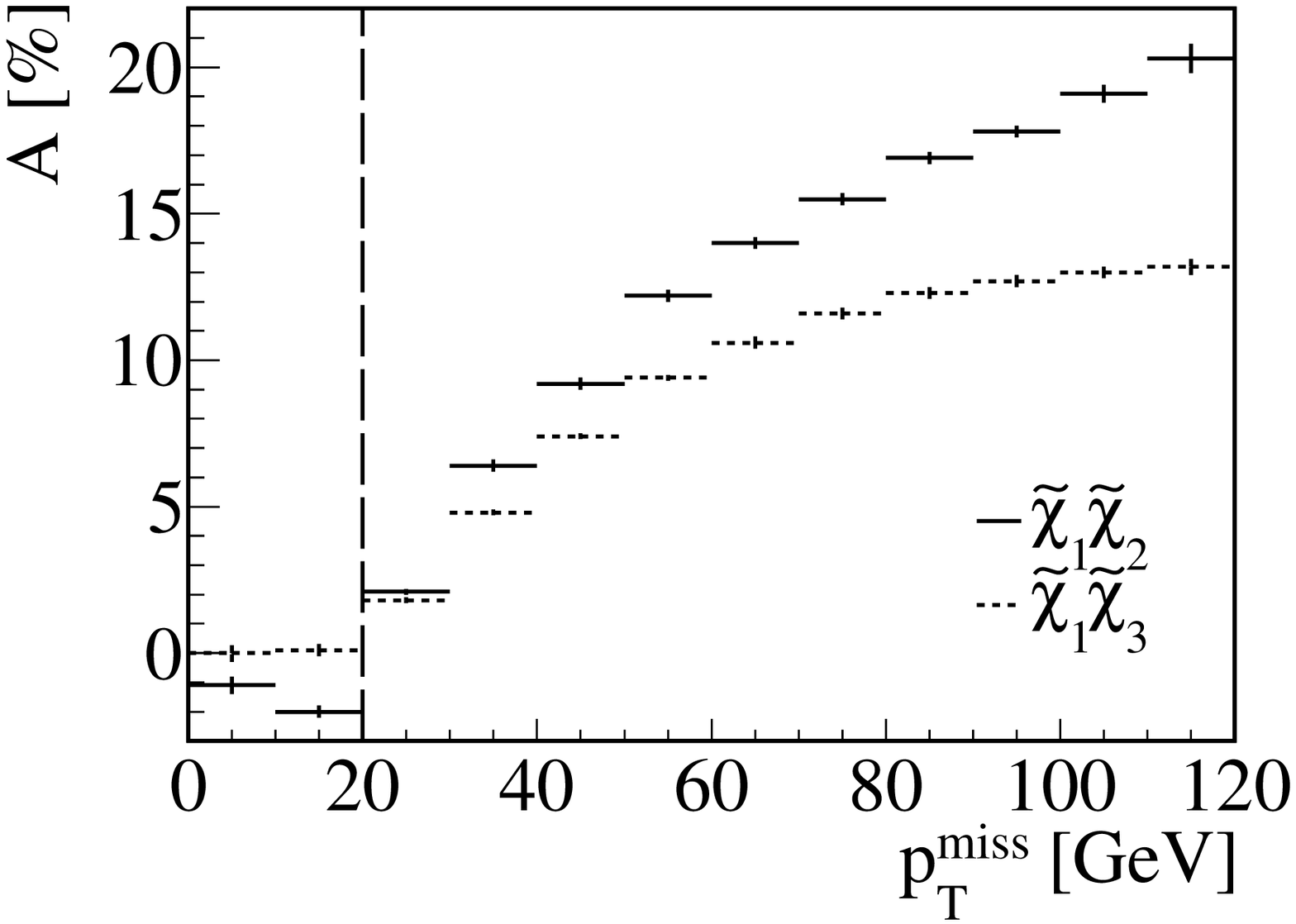}
  }
  \subfigure[]{\label{fig:acol_trip}
    \includegraphics[scale=0.35,angle=0]{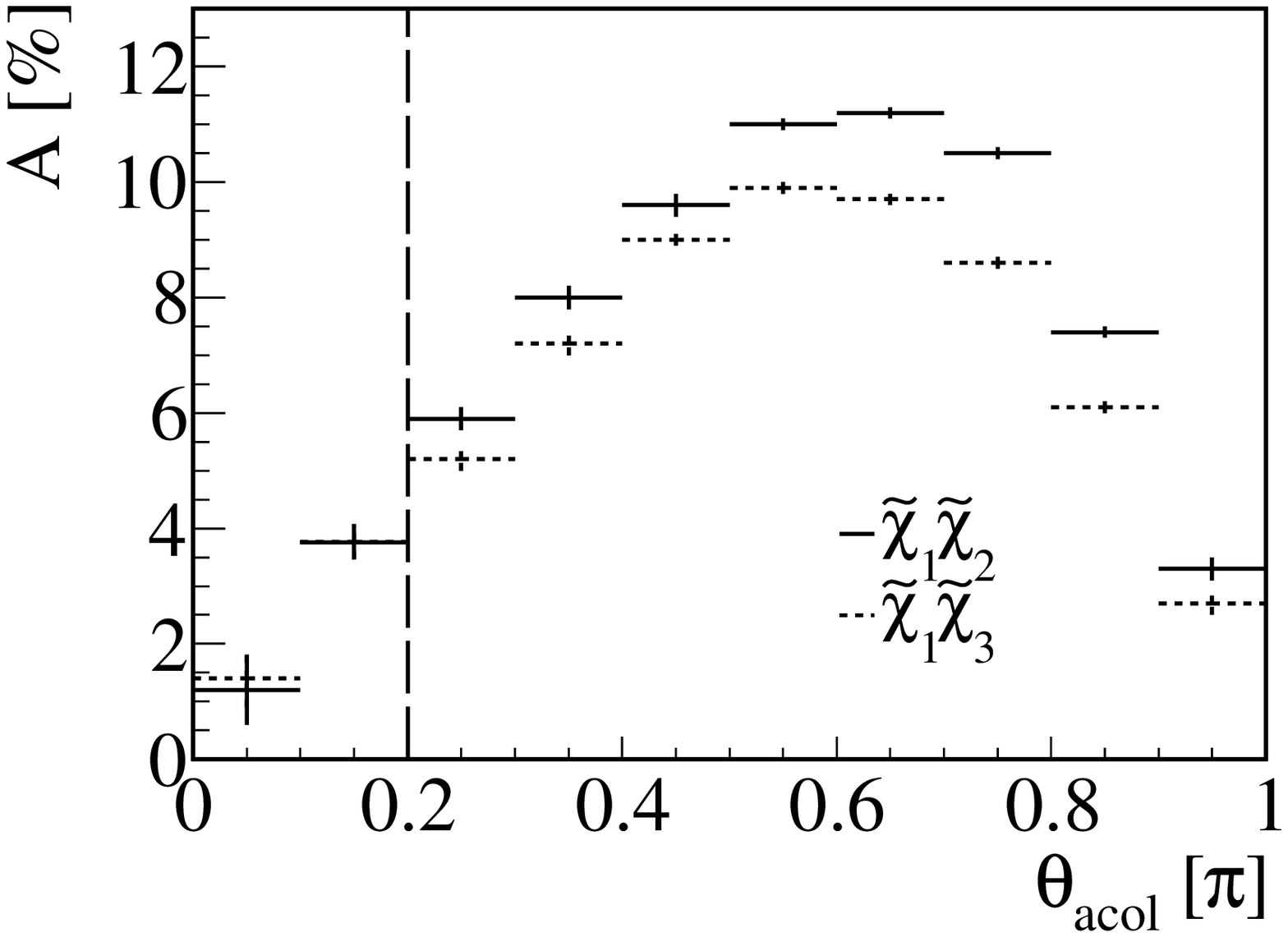}
  }
  \caption{ The (a) ${\mathbf p}_{\rm T}^{\rm miss}$ and (b)
    $\theta_{\rm acol}$ dependence of the asymmetries ${\mathcal
      A}({\mathbf p}_{e^-},{\mathbf p}_{\ell_N^-},{\mathbf
      p}_{\ell_F^+})_{\tilde\chi^0_1\tilde\chi^0_2}$ and ${\mathcal
      A}({\mathbf p}_{e^-},{\mathbf p}_{\ell_N^+},{\mathbf
      p}_{\ell_F^-})_{\tilde\chi^0_1\tilde\chi^0_3}$, solid and dashed
    lines, respectively. The cut value used in our analysis is
    indicated by the dashed line. In each case $10^7$ events were
    generated and no detector effects are included. Statistical
    uncertainties are shown. }
\label{fig:tripcuts}

\end{figure}

The CP asymmetries, Eq.~\eqref{eq:asyth}, can now be calculated as the
difference between the number of events $N_+$ and $N_-$, with the
triple product $\mathcal{T}>0$ or $\mathcal{T}<0$, respectively,
\begin{eqnarray}
{\mathcal A}(\mathcal{T}) &=& \frac{N_+ - N_-}{N_+ + N_-}.
\label{eq:asyex}
\end{eqnarray}
 We obtain 
\begin{eqnarray}
{\mathcal A}({\mathbf p}_{e^-},{\mathbf p}_{\ell_N^+},{\mathbf p}_{\ell_F^-})_{\tilde\chi^0_1\tilde\chi^0_2} &=& -10.2\pm 1.0\%, \\
{\mathcal A}({\mathbf p}_{e^-},{\mathbf p}_{\ell_N^-},{\mathbf p}_{\ell_F^+})_{\tilde\chi^0_1\tilde\chi^0_2} &=& +10.7\pm 1.0\%, \\
{\mathcal A}({\mathbf p}_{e^-},{\mathbf p}_{\ell_N^+},{\mathbf p}_{\ell_F^-})_{\tilde\chi^0_1\tilde\chi^0_3} &=& +9.3\pm 1.0\%, \\
{\mathcal A}({\mathbf p}_{e^-},{\mathbf p}_{\ell_N^-},{\mathbf p}_{\ell_F^+})_{\tilde\chi^0_1\tilde\chi^0_3} &=&  -8.8\pm 1.0\%,
\end{eqnarray}
with the statistical uncertainty~\cite{Desch:2006xp}
\begin{eqnarray}
   \delta(\mathcal{A})_{\mathrm{stat}} &=& \sqrt{\frac{1-\mathcal{A}^2}{N}},
\end{eqnarray}
and the total number of events $N = N_+ + N_-$. Exchanging the near
and far leptons gives, within the uncertainties, the same size of the
asymmetry but with opposite sign, see Eq.~\eqref{eq:leptonexch}.
However, the values of the asymmetries are different from the
theoretical values, see Tab.~\ref{tab:masses}, which is mainly due to:
\begin{enumerate}
 \item CP-even background events cancel in the numerator,
       but contribute to the denominator in Eq.~\eqref{eq:asyex}.
 \item Events are removed by the experimental selection cuts and
       by the kinematic selection procedure, which can bias the
       measured asymmetry.
\end{enumerate}
CP-even backgrounds shift the asymmetry to slightly lower values,
whereas the selection cuts have the opposite effect. If we assume that
the background contributions are known, we obtain
\begin{eqnarray}
{\mathcal A}({\mathbf p}_{e^-},{\mathbf p}_{\ell_N^+},{\mathbf p}_{\ell_F^-})_{\tilde\chi^0_1\tilde\chi^0_2} &=& -11.0 \pm 1.0\%,\label{asy:1}\\
{\mathcal A}({\mathbf p}_{e^-},{\mathbf p}_{\ell_N^-},{\mathbf p}_{\ell_F^+})_{\tilde\chi^0_1\tilde\chi^0_2} &=& +11.6 \pm 1.0\%,\label{asy:2}\\ 
{\mathcal A}({\mathbf p}_{e^-},{\mathbf p}_{\ell_N^+},{\mathbf p}_{\ell_F^-})_{\tilde\chi^0_1\tilde\chi^0_3} &=& +11.1 \pm 1.0\%,\label{asy:3}\\ 
{\mathcal A}({\mathbf p}_{e^-},{\mathbf p}_{\ell_N^-},{\mathbf p}_{\ell_F^+})_{\tilde\chi^0_1\tilde\chi^0_3} &=& -10.6 \pm 1.0\%.\label{asy:4}
\end{eqnarray}
Additionally, we have studied the bias due to the selection procedure
by calculating the asymmetry after each cut and comparing the results
obtained from \texttt{Whizard} with the results obtained from
\texttt{Herwig++}~\cite{Bahr:2008pv,Gieseke:2011na,Gigg:2007cr}. Both
programs show consistently a total shift of about 2\% towards larger
values after the application of the complete selection procedure; see
also Ref.~\cite{MoortgatPick:2010wp}. The cuts remove events with a
small value of the asymmetry inducing an upward shift. As an example,
Fig.~\ref{fig:tripcuts} shows the dependence of the asymmetry on
${\mathbf p}_{\rm T}^{\rm miss}$ and $\theta_{\rm acol}$, indicating
also our cut value. The shift will be taken into account in the
parameter fit described in the next section.

\section{Fit of the  parameters in the neutralino sector\label{fitting-proc}}

In the final step of our analysis, we estimate the accuracy to
determine the parameters in the neutralino sector of the MSSM. These
are the six free parameters of the neutralino mass matrix, see
App.~\ref{sec:NeutralinoMixing},
\begin{eqnarray} \label{eq:parameters}
|M_1|, \quad  M_2, \quad |\mu|,  \quad \tan\beta, \quad \phi_1, \quad\phi_\mu.
\end{eqnarray}
We have a number of observables at hand that can be used in the fit.
These are cross sections, masses, and asymmetries. Masses will be
measured with high precision using different methods~\cite{TDR}. For
the neutralino masses we assume the uncertainties as in
Ref.~\cite{Desch:2006xp}, since no detailed analysis has been done for
our parameter point. However, these uncertainties are rather
conservative, since we expect that in our scenario a similar precision
can be achieved as in~\cite{Martyn:2003av}. In case of the cross
sections, the uncertainty is dominated by the statistical uncertainty,
\begin{equation}
\frac{\Delta \sigma}{\sigma} = \frac{\sqrt{S+B}}{S},
\end{equation}
where $S$ and $B$ are the signal and background contributions,
respectively; see Tab.~\ref{tab:cutflow}. Since experimentally the
number of events is recorded, not the cross section itself, we have to
take into account branching ratios for the relevant decays. These will
depend on the masses and, in case of staus, on the stau mixing angle,
$\cos\theta_{\tilde{\tau}}$. The stau mixing angle can be obtained
from $\tau$ polarization measurements in stau pair
production~\cite{Bechtle:2009em}, with an accuracy of
5\%~\cite{Boos:2002wu,Boos:2003vf,Nojiri:1996fp}.

\medskip

After our procedure of the kinematic event selection, see
Sec.~\ref{sec:kinreco}, to disentangle contributions from
$\neu{1}\neu{2}$ and $\neu{1}\neu{3}$ production and decay, the
background contributions are below $15$\%. The small uncertainties in
the beam polarizations of $0.5$\%~\cite{MoortgatPick:2005cw}, in the
luminosity, and in the SUSY masses are negligible, see also
Ref.~\cite{ball,Desch:2003vw,D'Ascenzo:2009zz}. For the CP
asymmetries, we have estimated relative uncertainties of the order of
$10\%$ in Sec.~\ref{sec:asymmetryexp}. For the fit we take into
account a bias due to cuts on the asymmetry derived from the MC
simulation, as described in Sec.~\ref{sec:asymmetryexp}. Thus, the
analytical value of the asymmetry, given in Tab.~\ref{tab:masses}, is
shifted accordingly. Furthermore, we use Eq.~\eqref{eq:genuinecp} to
calculate the measured value of the asymmetry, which is free of FSI
effects. In summary, we have the following set of input observables
and uncertainties:
\begin{eqnarray*}
&& \mneu{1} = 117.3 \pm 0.2 \gev, \\
&& \mneu{2} = 168.5 \pm 0.5 \gev, \\
&& \mneu{3} = 180.8 \pm 0.5 \gev, \\
&& \sigma(\neu{1}\neu{2}) \times {\rm BR}(\neu{2}\to\tilde{\ell}_R\ell) = 130.9 \pm 1.4 \fb  ,\\
&& \sigma(\neu{1}\neu{3}) \times {\rm BR}(\neu{3}\to\tilde{\ell}_R\ell) =  155.7 \pm 1.6 \fb , \\
&& \sigma(\neu{2}\neu{2}) \times {\rm BR}(\neu{2}\to\tilde{\ell}_R\ell)^2  = 4.8 \pm 0.3 \fb , \\
&& \sigma(\neu{3}\neu{3}) \times {\rm BR}(\neu{3}\to\tilde{\ell}_R\ell)^2  = 26.3 \pm 0.7 \fb , \\
&& \sigma(\neu{2}\neu{3}) \times {\rm BR}(\neu{2}\to\tilde{\ell}_R\ell) \times {\rm BR}(\neu{3}\to\tilde{\ell}_R\ell)  = 28.9 \pm 0.7\fb ,   \\
&& {\mathcal{A}}^{\rm CP}({\mathbf p}_{e^-},{\mathbf p}_{\ell_N},{\mathbf p}_{\ell_F})_{\tilde\chi^0_1\tilde\chi^0_2} = +11.3\% \pm 0.7\%, \\
&& {\mathcal{A}}^{\rm CP}({\mathbf p}_{e^-},{\mathbf p}_{\ell_N},{\mathbf p}_{\ell_F})_{\tilde\chi^0_1\tilde\chi^0_3} = -10.9\% \pm 0.7\%.
\end{eqnarray*}
The uncertainties for the cross sections correspond to an integrated
luminosity of ${\mathcal{L}} = 500 \fb^{-1}$. We perform a six
dimensional $\chi^2$ fit using
\texttt{Minuit}~\cite{James:1975dr,minuit}
\begin{equation}
\chi^2 = \sum_i \left| \frac{{\mathcal{O}}_i - \bar{{\mathcal{O}}}_i }
                { \delta {\mathcal{O}}_i } \right|^2 ,
\end{equation}
where the sum runs over the input observables ${\mathcal{O}}_i$
mentioned above, with their corresponding experimental uncertainties
$\delta {\mathcal{O}}_i$. The theoretical values calculated using the
fitted MSSM parameters, Eq.~\eqref{eq:parameters}, are denoted by
$\bar{{\mathcal{O}}}_i$. The parameter dependence of branching ratios
(e.g. the stau mixing angle) is also included in the fit, but has
negligible impact. We then obtain the following fitted values for the
MSSM parameters:
\begin{eqnarray*}
 |M_1| &=& 150.0 \pm 0.7 \gev ,\\
  M_2 &=& 300 \pm 5 \gev ,\\
 |\mu| &=& 165.0 \pm 0.3 \gev ,\\
 \tan\beta &=& 10.0 \pm 1.6 , \\
 \phi_1 &=& 0.63 \pm 0.05 ,\\
 \phi_\mu &=& 0.0 \pm 0.2 .
\end{eqnarray*}
The best estimates are obtained for the $|M_1|$ and $|\mu|$ mass
parameters, since the neutralino states $\neu{1}$, $\neu{2}$, and
$\neu{3}$ are mostly composed of bino and Higgsino. The fourth
neutralino is heavy and cannot be measured, so the limit on the wino
mass $M_2$ is not as good. Also a rather large uncertainty is obtained
for $\tan\beta$. However, if additional measurements from other
sectors will be added, it should be improved significantly. We note
that the precision obtained in this study is similar to the results of
Ref.~\cite{Desch:2003vw}, which uses a similar set of
observables.\footnote {The high precision achieved in the fit calls
  for the inclusion of higher order corrections which can be in the
  $\mathcal{O}(20\%)$ regime in the neutralino system, see
  e.g.~\cite{Oller:2004br,Fritzsche:2004nf}. These corrections will in
  turn depend on the full parameter set of the MSSM. Therefore, the
  proper treatment would require the inclusion of observables from
  other sectors, in particular from the third generation of squarks,
  cf.~Ref.~\cite{Bechtle:2005vt,Lafaye:2007vs}.  This issue is beyond
  the scope of this paper, however, it should stimulate further
  studies.}

\medskip

It is remarkable that the moduli of the phases $\phi_1$ and $\phi_\mu$
can also be determined with high precision, using the CP-even
observables alone. However, only an inclusion of CP-odd asymmetries in
the fit allows us to resolve the sign ambiguities of the phases.
Without the CP-odd asymmetries in the fit we would have a twofold
ambiguity, $\phi_1 = \pm 0.6$, and even fourfold if $\phi_\mu \neq 0$.
Thus, the triple product asymmetries are not only a direct test of CP
violation, but are also essential to determine the correct values of
the phases.

\section{Summary and conclusions
	\label{Summary and conclusion}}

      We have presented the first full detector simulation study to
      measure SUSY CP phases at the ILC. We have considered
      CP-sensitive triple-product asymmetries in neutralino production
      $e^+e^- \to\tilde\chi^0_i \tilde\chi^0_1$ and the subsequent
      leptonic two-body decay chain $\tilde\chi^0_i \to \tilde\ell_R
      \ell$, $ \tilde\ell_R \to \tilde\chi^0_1 \ell$, for $ \ell=
      e,\mu$. Large asymmetries typically arise due to strong
      neutralino mixing. This causes on the one side that asymmetries
      for $\tilde\chi^0_1\tilde\chi^0_2$ and
      $\tilde\chi^0_1\tilde\chi^0_3$ production have about the same
      size but opposite sign. On the other side the strong mixing
      implies two close-in-mass neutralino states, that can have a
      mass separation of the order of $10$~GeV. This quasi-degeneracy
      would potentially pose a problem for the separation of both
      signal components.

\medskip

Therefore we have developed a kinematic selection method, to identify
the lepton pairs from the signal events. At the Monte Carlo
level, we have shown that this method allows one to separate the
leptons from the two signal processes $\tilde\chi^0_1\tilde\chi^0_2$
and $\tilde\chi^0_1\tilde\chi^0_3$, and also to reduce the major SM
and SUSY backgrounds, in particular from $W$-pair and slepton-pair
production.

\medskip

Then we have performed a detailed case study, which includes a full
ILD detector simulation and event reconstruction. A detailed cut flow
analysis has been done to preselect leptonic event candidates, which
then have been passed to our method of kinematic selection. Even after
the detector simulation, our method has worked efficiently to reduce
background and separate the signal.
After the full simulation with kinematic selection, the efficiencies
of signal event selection is of the order of $27\%$ with a purity of
about $90$\% of the event samples. That allows one to measure the
asymmetry with a relative precision of about $10\%$. {Our method of kinematic event reconstruction also works well in
scenarios with different mass splittings of the neutralinos and the
selectron. In the worst case scenario we found that the efficiency
will go down to some $10\%$, but still with a high purity of the
correctly identified signal sample of the order of $90\%$.} 

\medskip

We have performed a global fit of the neutralino masses, cross
sections, and CP asymmetries to reconstruct the MSSM parameters of the
neutralino sector, including the CP phases. The relative uncertainties
of the parameters $|M_1|$ and $|\mu|$ are below $1\%$, those for $M_2$
about $1\%$, and for $\tan\beta$ and the CP phases $\phi_1$,
$\phi_\mu$ about $10\%$. Although the moduli of the phases $\phi_1$,
$\phi_\mu$ can also be determined by using the CP-even observables
alone, we have shown that only an inclusion of CP-odd asymmetries in
the fit allows us to resolve the sign ambiguities of the phases.

\medskip

To summarize, we have shown that a measurement of the neutralino
sector seems to be feasible, including CP phases. In particular the
triple product asymmetries are not only a direct test of CP violation,
but are also essential to determine the correct values of the phases
in the neutralino sector.

\section*{Acknowledgments}

We would like to thank Steve Aplin, Mikael Berggren, Jan Engels, Frank
Gaede, Nina Herder, Jenny List, and Mark Thomson for useful discussions
and help with the detector simulations.
This work was supported by MICINN project FPA.2006-05294 and CPAN. We 
acknowledge the support of the DFG through the SFB (grant SFB 676/1-2006).

\begin{appendix}

\renewcommand{\thesubsection}{\Alph{section}.\arabic{subsection}}
\renewcommand{\theequation}{\Alph{section}.\arabic{equation}}

\setcounter{equation}{0}

\section{Reconstruction of $W$ and $\tilde\ell$ pair production}\label{sec:app3}

We consider a template process
\begin{equation}
e^+ + e^- \;\to\; A + \bar A \;\to\; \ell + \bar\ell + B + B,
\end{equation}
where $(A,B) = (\tilde\ell,\neu{1})$ or $(W,\nu)$. In both cases
$B=\neu{1}$ or $\nu$ escapes detection. Since the system of the lepton
pair has to obey different kinematic constraints, we consider the
question, whether the final lepton pair can be assigned to its mother
production process, if the lepton momenta are measured, and the
slepton and LSP masses are known.  We follow closely
Ref.~\cite{Buckley:2007th}, and define the notation
\begin{eqnarray}
c_1 &\equiv& {\mathbf p}_A \cdot {\mathbf p}_\ell \;= \;\phantom{-}
       \frac{1}{2} ( m^2_B -m_A^2 +
                     E_\ell \sqrt{s}),\label{eq:c1}\\
c_2 &\equiv& {\mathbf p}_A \cdot {\mathbf p}_{\bar\ell} \;=\; 
        -\frac{1}{2} ( m^2_B - m^2_A +  E_{\bar\ell} \sqrt{s}),
   \label{eq:c2}\\
b_2 &\equiv& {\mathbf p}_A \cdot {\mathbf p}_A \;=\; \frac{s}{4} - m^2_A,
     \label{eq:b2} \\[2mm]
a_{11} &\equiv& {\mathbf p}_\ell \cdot {\mathbf p}_\ell, \quad
a_{12} \;\equiv\; {\mathbf p}_\ell \cdot {\mathbf p}_{\bar\ell}, \quad
a_{22} \;\equiv\; {\mathbf p}_{\bar\ell} \cdot {\mathbf p}_{\bar\ell}.
\end{eqnarray}
The momentum ${\mathbf p}_A$ can be decomposed into the final lepton momenta
\begin{eqnarray}
{\mathbf p}_A &= & t_1 \, {\mathbf p}_\ell + t_2\,  {\mathbf p}_{\bar\ell} + 
                   y \, {\mathbf p}_\bot,
\end{eqnarray}
where ${\mathbf p}_\bot = {\mathbf p}_\ell \times {\mathbf p}_{\bar\ell}$. 
The expansion coefficients follow from Eqs.~\eqref{eq:c1} and \eqref{eq:c2}
\begin{eqnarray}
\left|
\begin{array}{ccc}
c_1 &=& t_1 a_{11} + t_2 a_{12} \\[3mm]
c_2 &=& t_1 a_{12} + t_2 a_{22}
\end{array}
\right|
&\Rightarrow&
\left|
\begin{array}{ccc}
t_1 &=& \displaystyle{\frac{a_{22} c_1 - a_{12} c_2}{a_{11}a_{22} - a_{12}^2}} \\[5mm]
t_1 &=& \displaystyle{\frac{a_{11} c_2 - a_{12} c_1}{a_{11}a_{22} - a_{12}^2}}
\end{array}
\right|.
\label{eq:ct12} 
\end{eqnarray}
We finally obtain,  from Eqs.~\eqref{eq:b2} and \eqref{eq:ct12},
\begin{eqnarray}
  b_2 &=& (t_1^2 a_{11} + 2 t_1 t_2 a_{12} + t_2^2 a_{22}) + y^2
  |{\mathbf p}_\bot|^2, 
  \qquad\\[4mm]
  \Rightarrow
  y^2
  &=& \frac{b_2 - (t_1^2 a_{11} + 2 t_1 t_2 a_{12} + t_2^2
    a_{22})}{|{\mathbf p}_\bot|^2}.\label{eq:defy}
\end{eqnarray}
The equation for $y$ constitutes a condition for existence of physical
solutions of the system, i.e.\ $y^2 \geq 0$, where $y^2$ is computed
from the kinematic variables $s$, $m_A$, $m_B$, $E_\ell$,
$E_{\bar\ell}$, and ${\mathbf p}_\ell \cdot {\mathbf p}_{\bar\ell}$.
Similar to neutralino pair production, Sec.~\ref{sec:neurec},
Eq.~\eqref{chipm}, there remains a twofold ambiguity in solving the
$W$ or $\tilde{\ell}$ system, $y = \pm \sqrt{y^2}$.

\setcounter{equation}{0}
\section{Neutralino mixing}
\label{sec:NeutralinoMixing}

The complex symmetric mass matrix of the neutralinos in the photino, 
zino, Higgsino basis ($\tilde{\gamma},\tilde{Z}, \tilde{H}^0_a, \tilde{H}^0_b$), 
is given by~\cite{Choi:2001ww}
\begin{equation}
	{\mathcal M}_{\chi^0} =
	\left(\begin{array}{cccc}
		M_2 \, s^2_W + M_1 \, c^2_W & 
		(M_2-M_1) \, s_W c_W & 0 & 0 \\
		(M_2-M_1) \, s_W c_W & 
		M_2 \, c^2_W + M_1 \, s^2_W & m_Z  & 0 \\
		0 & m_Z &  \mu \, s_{2\beta} & -\mu \, c_{2\beta} \\
		0 &  0  & -\mu \, c_{2\beta} & -\mu \, s_{2\beta} 
	\end{array}\right),
\label{eq:neutmass}
\end{equation}
with the short hand notation for the angles $s_W = \sin\theta_W$, $c_W
= \cos\theta_W$, and $s_{2\beta} = \sin(2\beta)$, $c_{2\beta} =
\cos(2\beta)$, and the $SU(2)$ gaugino mass parameter $M_2$. The
phases of the complex parameters $M_1=|M_1|e^{i\phi_{1} }$ and
$\mu=|\mu|e^{i\phi_\mu } $ can lead to CP-violating effects in the
neutralino system. The phase of $M_2$ can be rotated away by a
suitable redefinition of the fields. We diagonalize the neutralino
mass matrix with a complex, unitary $4\times 4$ matrix $N$,
\begin{equation}
	N^* \cdot {\mathcal M}_{\chi^0} \cdot N^{\dagger} =
	{\rm diag}(m_{\chi^0_1},\dots,m_{\chi^0_4}),
\label{eq:neutn}
\end{equation}
with the real neutralino masses $ 0 < m_{\chi^0_1} < m_{\chi^0_2} <
m_{\chi^0_3} < m_{\chi^0_4}$.

\end{appendix}

\bibliographystyle{utphys}
\bibliography{cp_ilc}

\end{document}